\documentclass[11pt]{article}
\usepackage{subfigure}
\usepackage{graphicx}
\usepackage{natbib}
\usepackage{amsfonts}
\usepackage{amssymb}
\usepackage{amsmath}
\usepackage{lscape}
\usepackage{color}
\usepackage[left=1.7cm, right=2.5cm, top=1.7cm, bottom=1.7cm]{geometry}
\usepackage{algorithm}
\usepackage{algorithmic}

\definecolor{DarkO}{cmyk}{0.0, 0.43, 0.90, 0.13}

\def\bt{\boldsymbol{\theta }}
\DeclareMathOperator*{\argmax}{arg\,max}
\DeclareMathOperator*{\argmin}{arg\,min}

\begin{document}

\title{Approximating Bayes in the 21st Century\thanks{{\footnotesize The
authors would like to thank the Editor, an associate editor and two
anonymous reviewers for very constructive and insightful comments on an
earlier draft of the paper. That earlier version, comprising a more
extensive coverage of Bayesian computation \textit{per se}, and an
historical timeline for computational developments, appears under the title
`Computing Bayes: Bayesian Computation from 1763 to the 21st Century' (%
\citealp{martin2020computing}). Martin and Frazier have been supported by
Australian Research Council Discovery Grants DP170100729 and DP200101414,
and the Australian Centre of Excellence in Mathematics and Statistics.
Frazier has also been supported by Australian Research Council Discovery
Early Career Researcher Award DE200101070. Robert has been partly supported
by a senior chair (2016-2021) from l'Institut Universitaire de France and by
a Prairie chair from the Agence Nationale de la Recherche
(ANR-19-P3IA-0001). }}}
\author{Gael M. Martin\thanks{{\footnotesize Corresponding author:
gael.martin@monash.edu.}}, David T. Frazier and Christian P. Robert}
\maketitle

\begin{abstract}
The 21st century has seen an enormous growth in the development and use of
approximate Bayesian methods. Such methods produce computational solutions
to certain `intractable' statistical problems that challenge exact methods
like Markov chain Monte Carlo: for instance, models with unavailable
likelihoods, high-dimensional models, and models featuring\textbf{\ }large
data sets. These approximate methods are the subject of this review. The aim
is to help new researchers in particular -- and more generally those
interested in adopting a Bayesian approach to empirical work -- distinguish
between{\ different approximate techniques; understand the sense in which
they are approximate; appreciate when and why particular methods are useful;
and see }the ways in which they can can be combined.

\bigskip


\bigskip

\emph{MSC2010 Subject Classification}: 62-03, 62F15, 65C60
\end{abstract}

\newpage

\section{Introduction}

\baselineskip17.5pt

The advent of fast, accessible computers in the last two decades of the 20th
century (\citealp{ceruzzi:2003}), allied with the exploitation of earlier
insights into probabilistic simulation (\citealp{metropolis:ulam:1949}; %
\citealp{metropolis:1953}; \citealp{hammersley:handscomb:1964}; %
\citealp{hastings:1970}), led to an explosion in the use of simulation-based
computation to solve empirical Bayesian problems. Whilst significant
advances were made in econometrics (\citealp{kloek1978bayesian}; %
\citealp{bauwens:1985}; \citealp{geweke:1989}) and signal processing %
\citep{gordon:salmon:smith:1993} using the principles of importance
sampling, the `computational revolution' -- as it is often coined -- was
driven primarily by Markov chain Monte Carlo (MCMC) algorithms; see \cite%
{geman:1984}, \cite{tanner87} and \cite{gelfand:smith90} for seminal
contributions, and \cite{besag:green:1993}, \cite{smith:roberts:1993}, 
\cite{chib2011introduction}%
, 
\cite{geyer2011introduction}
and \cite{robert:casella:2011} for selected reviews.

The impact {of these} computational advances was felt across a huge array of
fields -- genetics, biology, neuroscience, astrophysics, image analysis,
ecology, epidemiology, engineering, education, economics, political science,
marketing and finance, to name but some -- and brought Bayesian analysis
into the statistical mainstream. The three handbooks: \textit{The Oxford
Handbook of Applied Bayesian Analysis} (\citealp{ohagan2010handbook}), 
\textit{Handbook of Markov Chain Monte Carlo }(\citealp{brooks:etal:2011})\
and \textit{The Oxford Handbook of Bayesian Econometrics} (%
\citealp{geweke2011handbook}), highlight the wide spectrum of fields, and
broad scope of empirical problems to which MCMC and importance sampling
algorithms were (and continue to be)\textbf{\ }applied; as do certain
contributions to the series of vignettes edited by Mengersen and Robert for 
\textit{Statistical Science} (2014, Vol 29, No. 1), under the theme of `Big
Bayes Stories'.

Despite their unquestioned power and versatility however, these original
simulation techniques did have certain limitations; with these limitations
to become more marked as the empirical problems being tackled became more
ambitious; and this despite a concurrent rise in computing power (parallel
computing, access to graphical processing units, and so forth). In short,
the early algorithms were to stumble in the face of so-called `intractable'
statistical problems: data generating processes with likelihoods\ that are
unavailable analytically; models with a very large number of unknowns; and
models featuring\textbf{\ }`big data'. `Exact' solutions to such problems
were simply not achievable via the early {MCMC and importance sampling
algorithms} or, at least, were not available in a {reasonable} computing
time; and `approximate' solutions were, instead, often sought. It is those
approximate solutions that are the subject of this review.\footnote{%
We acknowledge, of course, that there have been \textit{many} concurrent
advances in MCMC and importance sampling designed, in particular, to deal
with the problem of scale. We refer the reader to: 
\cite{greenetal2015}, \cite{robert2018accelerating} and \cite%
{dunson2019hastings} for broad overviews of modern developments in MCMC; {to 
\cite{betancourt:2018} for a review of Hamiltonian Monte Carlo (HMC); }to 
\cite{naesseth2019elements} for a recent review of sequential Monte Carlo
(exploiting importance sampling principles as it does); and to \cite%
{hoogerheide2009simulation}, \cite{Tokdar2010} and \cite{elvira2021advances}
for other advances in importance sampling.}

Our overarching aim is to provide readers with some insight into questions
such as: `{In what sense are approximate methods of computation} \textit{%
`approximate}'?', `What are the connections between different approximate
methods?', `When does one use one approach, and when another?', and `Can
different methods be combined to tackle multiple, distinct instances of
`intractability'?'. In order to address such questions, we bring together in
one place, and using a common notational framework, the four main
approximate techniques that have evolved during the 21st century:
approximate Bayesian computation (ABC), Bayesian synthetic likelihood (BSL),
variational Bayes (VB) and integrated nested Laplace approximation\textit{\ }%
(INLA). One goal is to link the development of these new techniques to the
increased complexity, and size, of the empirical problems being analyzed. A
second goal is to draw out insightful links \textit{and} differences between
all and, in so doing, pinpoint when and why each technique has value, with
illustrative examples from the literature used to enhance this
demonstration. This {then }provides some context for the \textit{hybrid}
computational methods that we then review. Whilst formally providing an
exact solution and, hence, not a focus of this paper, we give a brief
outline of pseudo-marginal methods, including particle MCMC (PMCMC), due to
the role such methods play -- in tandem with certain approximate techniques
-- under the `hybrid' umbrella.

{This paper is meant to serve both as a `first port of call' for those who
are new to modern Bayesian computation, and as a\ useful overview for
practitioners with established, but selective, expertise.} {Hence, e}%
xcessive formalism, and extensive {algorithmic }detail, is avoided in order
to make the paper as accessible as possible, {and to keep the focus on the }%
key principles{\ underpinning each computational method. We do not attempt
to replicate the coverage of existing reviews of specific approximate
methods. Rather, we direct readers to those review papers and handbook
chapters where necessary, including for coverage of all published work}.\
Whilst we make brief reference to various software packages, we also defer
to those other resources for detailed descriptions of the dedicated software that is available for
implementing {particular computational techniques}.

The remainder of the paper is as follows. In Section \ref{general} we
provide a brief outline of the general Bayesian computational problem, and
explain when that problem may be viewed as intractable. Section \ref{approx}
then outlines the approximate solutions to such intractable problems, ABC,
BSL, VB and INLA, and hybridized versions thereof. {For ease of exposition
-- and with acknowledgement that this categorization} is imperfect -- these
main methods{\ are grouped into: `Simulation-based Approaches' (ABC and BSL)
and `Optimization Approaches' (VB and INLA). Several hybrid approximate
methods are then described, in which distinct }methods are amalgamated for
the purpose of simultaneously solving multiple computational challenges{\
(e.g., a high-dimensional} model with a computationally expensive, or
analytically unavailable{\ likelihood).} In order to illustrate the type of
intractable problems for which approximate solutions have been sought, we
display the results of\ {selected empirical illustrations} from the
literature in which, respectively, {simulation-based computation} and {%
optimization-based computation} have been used. {The paper concludes in
Section \ref{future}} {with some perspectives on the future. Particular
attention is given to three directions that the authors believe to be worthy
of attention: \textit{1) }The performance of approximate methods under
(model) misspecification; \textit{2) }The use of approximate methods in
generalized\textit{\ }(non-likelihood) settings; and, finally, \textit{3) }%
The role to be played by approximate inference in Bayesian prediction.}

\section{Bayesian Computation in a Nutshell \label{general}}

\subsection{A short primer\label{primer}}

We being by establishing notation. An $n$-dimensional vector of observed
data $\mathbf{y}=(y_{1},y_{2},...,y_{n})^{\prime }$ is assumed to be
generated from some data generating process (DGP) $p(\mathbf{y}|\boldsymbol{%
\theta })$, with $\boldsymbol{\theta }\in {\Theta \subseteq }$ $\mathbb{R}%
^{p}$ a $p$-dimension vector of unknown parameters, and where we possess
prior beliefs on $\boldsymbol{\theta }$ specified by the prior probability
density function (pdf) $p(\boldsymbol{\theta })$. By Bayes' rule, the joint
posterior pdf $p(\boldsymbol{\theta }\mathbf{|y})$ is defined by 
\begin{equation}
p(\boldsymbol{\theta }\mathbf{|y})\propto p(\mathbf{y|}\boldsymbol{\theta }%
)p(\boldsymbol{\theta }).  \label{exact_posterior}
\end{equation}%
Most Bayesian quantities of interest are, in turn, posterior\textit{\ }%
expectations of some function $g(\boldsymbol{\theta })$ and, hence, can be
expressed as,

\begin{equation}
\mathbb{E}(g(\boldsymbol{\theta }\mathbf{)|y)}=\int_{{\Theta }}g(\boldsymbol{%
\theta }\mathbf{)}p(\boldsymbol{\theta }\mathbf{|y})d\boldsymbol{\theta }%
\mathbf{.}  \label{gen_expect}
\end{equation}%
Familiar examples include posterior moments, such as 
\begin{equation*}
\mathbb{E}(\boldsymbol{\theta }\mathbf{|y)}=\int_{{\Theta }}\boldsymbol{%
\theta }p(\boldsymbol{\theta }\mathbf{|y})d\boldsymbol{\theta }\text{ and Var%
}(\boldsymbol{\theta }\mathbf{|y)}=\int_{{\Theta }}\left[ \boldsymbol{\theta 
}-\mathbb{E}(\boldsymbol{\theta }\mathbf{|y)}\right] \left[ \boldsymbol{%
\theta }-\mathbb{E}(\boldsymbol{\theta }\mathbf{|y)}\right] ^{\prime }p(%
\boldsymbol{\theta }\mathbf{|y})d\boldsymbol{\theta }\mathbf{,}
\end{equation*}%
plus marginal quantities like $p(\theta _{1}^{\ast }|\mathbf{y})=\int_{{%
\Theta }}p(\theta _{1}^{\ast }|\theta _{2},...,\theta _{p},\mathbf{y})p(%
\boldsymbol{\theta }\mathbf{|y})d\boldsymbol{\theta }$ (for $\theta
_{1}^{\ast }$ a point in the support of $p(\theta _{1}|\mathbf{y})$).
However, (\ref{gen_expect}) also subsumes the case where $g(\boldsymbol{%
\theta }\mathbf{)}=$ $p(y_{n+1}^{\ast }|\boldsymbol{\theta }\mathbf{,y})$
(with $y_{n+1}^{\ast }$\textbf{\ }in the support of the `out-of-sample'
random variable, $y_{n+1}$), in which case (\ref{gen_expect}) defines the 
\textit{predictive} distribution for $y_{n+1}$:%
\begin{equation}
p(y_{n+1}^{\ast }|\mathbf{y})=\int_{{\Theta }}p(y_{n+1}^{\ast }|\boldsymbol{%
\theta }\mathbf{,y})p(\boldsymbol{\theta }\mathbf{|y})d\boldsymbol{\theta }.
\label{predict}
\end{equation}%
It also encompasses $g(\boldsymbol{\theta }\mathbf{)}=$ $L(\boldsymbol{%
\theta },d)$,\textbf{\ }for $L(\boldsymbol{\theta },d)$ a loss function
associated with a decision $d$, in which case (\ref{gen_expect}) is the
quantity minimized in Bayesian decision theory (\citealp{berger:1985}; %
\citealp{robert:2001}). Further, defining $g(\boldsymbol{\theta }\mathbf{)}=$
$p(\mathbf{y}|\boldsymbol{\theta },\mathcal{M})$ as the DGP that explicitly
conditions on a model, $\mathcal{M}$ say, the marginal likelihood of $%
\mathcal{M},$ $p(\mathbf{y}|\mathcal{M})$, is the expectation,%
\begin{equation}
\mathbb{E}(g(\boldsymbol{\theta }\mathbf{)|}\mathcal{M}\mathbf{)}=\int_{{%
\Theta }}g(\boldsymbol{\theta }\mathbf{)}p(\boldsymbol{\theta }|\mathcal{M})d%
\boldsymbol{\theta },  \label{gen_expect_prior}
\end{equation}%
with respect to the\ prior, $p(\boldsymbol{\theta }|\mathcal{M})$. The ratio
of (\ref{gen_expect_prior}) to the comparable quantity for an alternative
model $\mathcal{M}^{\prime }$ defines the Bayes factor for use in choosing
between the two models. In summary then, the key quantities that underpin
the whole of Bayesian analysis -- inference, prediction, decision theory,
and model choice -- can be expressed as expectations.

The need for numerical computation arises simply because \textit{analytical}
solutions to (\ref{gen_expect}) and (\ref{gen_expect_prior}) are rare.
Typically, the posterior does not possess a closed form, as the move from
the generative problem (the specification of $p(\mathbf{y}|\boldsymbol{%
\theta })$) to the inverse problem (the production of $p(\boldsymbol{\theta }%
\mathbf{|y})$), yields a posterior that is known only up to the constant of
proportionality. The availability of $p(\boldsymbol{\theta }\mathbf{|y})$
only up to the integrating constant\textit{\ }immediately{\ precludes} the
analytical solution of (\ref{gen_expect}), for any $g(\boldsymbol{\theta })$%
. By definition, a lack of knowledge of the integrating constant implies
that the marginal likelihood for the model in (\ref{gen_expect_prior}) is
unavailable. Hence the need for computational\textit{\ }solutions.

It is useful to think about all Bayesian computational techniques falling
into one of three broad categories: \textit{1) Deterministic integration
methods; 2) Exact simulation methods; 3) Approximate (including asymptotic)
methods. }Whilst all techniques are applicable to both the posterior
expectation in (\ref{gen_expect}) and the prior expectation in (\ref%
{gen_expect_prior}), to keep the scope of the paper manageable we only
consider the computation of (\ref{gen_expect}).\footnote{%
See \cite{ARDIA2012} and \cite{llorente2021marginal} for extensive reviews
of methods for computing marginal likelihoods.} In brief, the methods in 
\textit{1) }define $L$ grid-points, $\boldsymbol{\theta }_{1},$ $\boldsymbol{%
\theta }_{2},...,\boldsymbol{\theta }_{L}$, to span the support of $%
\boldsymbol{\theta }$, compute $g(\boldsymbol{\theta }_{l}\mathbf{)}p(%
\boldsymbol{\theta }_{l}\mathbf{|y})$, for $l=1,2,...,L$, and estimate (\ref%
{gen_expect}) as a weighted sum of these $L$ values of the integrand.
Different deterministic numerical integration (or quadrature) rules are
based on different choices for $\boldsymbol{\theta }_{l}$, $l=1,2,...,L$,
and different formulae for the weights (see \citealp{davis1975numerical}, %
\citealp{naylor:smith:1982}, \citealp{vanslette2019simple}, and %
\citealp{bilodeau2021stochastic}, for relevant coverage). Such methods
remain an important tool in the Bayesian arsenal, and we note the recent
explosion of probabilistic numerics creating new connections between
Bayesian concepts and numerical integration (Briol \textit{et al}., 2019).
However, deterministic integration -- on its own -- plays a relatively small
role in Bayesian numerical work due primarily to the `curse of
dimensionality' from which it suffers.

The methods in \textit{2) }use simulation\textit{\ }to produce $M$ draws of $%
\boldsymbol{\theta }$, $\boldsymbol{\theta }^{(i)}$, \thinspace $i=1,2,...,M$%
, from $p(\boldsymbol{\theta }\mathbf{|y})$, with the mean of the $M$
transformed draws, $g(\boldsymbol{\theta }^{(i)}\mathbf{)}$, often used to
estimate (\ref{gen_expect}). Different simulation methods are distinguished
by the way in which the draws are produced, and whether those draws are
independent (e.g. Monte Carlo simulation; importance sampling) or dependent
(e.g., MCMC). However, under appropriate regularity, like finite variance,
and subject to convergence in the case of MCMC, all such methods produce a $%
\sqrt{M}$-consistent estimate of (\ref{gen_expect}), whatever the degree of
dependence in the draws, with the dependence affecting the constant implicit
in the $O(M^{-1/2})$ term, but not the rate itself (%
\citealp{geyer2011introduction}). Hence, in principle, any algorithm that
simulates from $p(\boldsymbol{\theta }\mathbf{|y})$ can produce an estimate
of (\ref{gen_expect}) that is arbitrarily accurate for large enough $M;$
justifying the use of the adjective `exact'.

Finally, the methods in \textit{3) }replace the integrand in (\ref%
{gen_expect}) with an approximation of some sort, and evaluating the
resultant integral. Different approximation methods are defined by the
choice of replacement for the integrand, with the nature of this replacement
determining the way in which the final integral is computed. \textit{%
Asymptotic }approximation methods replace the integrand with an expansion
that is accurate for large $n$, and yield an estimate of (\ref{gen_expect})
that is accurate asymptotically.

{It is the class of approximate methods in \textit{3) }that is our focus,
with our first task} being to establish why \textit{these }methods, rather
than those in \textit{2)},\textit{\ }are useful in intractable settings.

\subsection{Intractable Bayesian problems\label{intract}}

With reference to (\ref{exact_posterior}), two characteristics are worthy of
note. First, as is common knowledge, in all but the most stylized problems
(for example, when $p(\mathbf{y}|\boldsymbol{\theta })$ is from the
exponential family, and either a natural conjugate, or convenient
noninformative prior is adopted), $p(\boldsymbol{\theta }\mathbf{|y})$ is
available only\textit{\ }up to its integrating constant, and cannot be
directly simulated. Second, representation of $p(\boldsymbol{\theta }\mathbf{%
|y})$ only as a kernel, $p^{\ast }(\boldsymbol{\theta }\mathbf{|y})\propto p(%
\mathbf{y}|\boldsymbol{\theta })p(\boldsymbol{\theta })$, still requires
closed forms for $p(\mathbf{y}|\boldsymbol{\theta })$ and $p(\boldsymbol{%
\theta }).$ With reference to $p(\mathbf{y}|\boldsymbol{\theta })$, this
means that, for any $\boldsymbol{\theta }$, $p(\mathbf{y}|\boldsymbol{\theta 
})$ needs to be able to be evaluated at the observed $\mathbf{y}$. The MCMC
and importance sampling methods obviate the first problem by drawing \textit{%
indirectly }from $p(\boldsymbol{\theta }\mathbf{|y})$ via another
(`candidate' or `proposal') distribution from which simulation is feasible.
However, these methods still require evaluation of $p(\mathbf{y}|\boldsymbol{%
\theta })$: in the computation of the importance weights in importance
sampling (\citealp{geweke:1989}; \citealp{Tokdar2010}), in the computation
of the acceptance probability in any Metropolis-Hastings MCMC algorithm (%
\citealp{hastings:1970}; \citealp{chib:greenberg:1995}), and in the
implementation of any Gibbs-based MCMC algorithm, in which the conditional
posteriors are required either in full form or at least up to a scale factor
(\citealp{casella:george:1992}; \citealp{chib_greenberg_1996}).\footnote{%
Some versions of these methods only require a term of $p(\mathbf{y}|%
\boldsymbol{\theta })$ to be available or allow for its replacement by an
unbiased estimate, as in pseudo-marginal MCMC; see Section \ref{sec:psudo}.}

The assumption that $p(\mathbf{y}|\boldsymbol{\theta })$ can be evaluated is
a limitation for two reasons. First, some DGPs do not admit pdfs ({or
probability mass functions) }in closed form; examples being: probability
distributions defined by quantile or generating functions (%
\citealp{devroye:1986}; \citealp{peters2012likelihood}), continuous time
models in finance with unknown transition densities %
\citep{gallant1996moments}, dynamic equilibrium models in economics %
\citep{calvet2015accurate}, certain deep learning models in machine learning %
\citep{goodfellow2014generative}; complex astrophysical models %
\citep{jennings2017astroabc}; and\ DGPs for which the normalizing constant
is unavailable, such as Markov random fields in spatial modelling (%
\citealp{rue:held:2005}; \citealp{stoehr2017review}). Second, pointwise
evaluation of $p(\mathbf{y}|\boldsymbol{\theta })$ (at any $\boldsymbol{%
\theta }$) (in the case where $p(\boldsymbol{\cdot }|\boldsymbol{\theta })$ 
\textit{has} a closed form) entails an $O(n)$ computational burden; meaning
that the original MCMC and importance sampling methods are not readily%
\textit{\ scalable} to so-called `big (or tall) data' problems %
\citep{bardenet2017markov}.

Just as important are the challenges that arise when the dimension of the
unknowns themselves is very large (the so-called `high-dimensional'
problem). A prime example of this is when the vector of unknowns, $%
\boldsymbol{\theta }$, comprises both a set of fixed\textbf{\ }`global'
parameters that govern all the data (call this set $\boldsymbol{\phi }$),
and a set of latent random variables that are `local' to individual data
points (call this set $\mathbf{x}$), and where $\mathbf{x}$ is very large --
sometimes of dimension exceeding $n$ (e.g. %
\citealp{tavare:balding:griffith:donnelly:1997}; {%
\citealp{rue:martino:chopin:2009}}; \citealp{beaumont:2010}; %
\citealp{braun2010variational}; \citealp{lintusaari2017fundamentals}; %
\citealp{johndrow2019mcmc}). In such cases, standard MCMC methods -- even if
feasible in principle -- may not enable an accurate estimate of (\ref%
{gen_expect}) to be produced in finite computing time;\ i.e. such methods
are \textit{not necessarily scalable} in the dimension of the unknowns (%
\citealp{betancourt:2018}). {We alert the reader to the fact that the term
`intractable likelihood' is sometimes used to refer to such cases, }since
the likelihood for the global parameters,{\textbf{\ }}$p(\mathbf{y}|%
\boldsymbol{\phi })${\textbf{\ -- }}which requires integration over the
latent parameters -- is typically not available in closed form, even when
the `complete' likelihood,{\ }$p(\mathbf{y,x}|\boldsymbol{\phi })=p(\mathbf{y%
}|\mathbf{x,}\boldsymbol{\phi })p(\mathbf{x|}\boldsymbol{\phi })$,{\ }%
\textit{is}\textbf{\ }available.\textbf{\ }We reserve the term `intractable'
or `unavailable' likelihood for the case where the DGP cannot be expressed
in closed form. The intractability that arises in problems with a large
number of latent variables we view as simply an example of the computational
difficulties that arise when\textbf{\ }$\boldsymbol{\theta }$ is of high
dimension.

The techniques discussed in Section \ref{approx} relieve the investigator of
one or more of these instances of intractability: unavailable likelihood;
high-dimensional $\boldsymbol{\theta };$ `big' $\mathbf{y}$. But they come
at a cost. All such methods produce an estimate of (\ref{gen_expect}) that
is only ever intrinsically approximate.

\section{Approximate Bayesian Methods\label{approx}}

As noted above, the goal of all exact simulation-based computational methods
(including the pseudo-marginal techniques that play a role in the hybrid
approximation methods discussed in Section \ref{hybrid}), is to estimate the
posterior expectation in (\ref{gen_expect}) `exactly', at least up to some $%
O(M^{-1/2})$ term, where $M$ is the number of draws that defines the
simulation scheme. The alternative methods do, of course, differ one from
the other in terms of the constant term that quantifies the precise error of
approximation. Hence, it may be the case that even for a very large $M$, a
nominally exact method (despite being `tuned' optimally) has an
approximation error that is non-negligible. Nevertheless, the convention in
the literature is to refer to all simulation methods outlined to this point
as \textit{exact}, typically without qualification.\footnote{%
We note that so-called `quasi-Monte Carlo' integration schemes aim for
exactness at a faster rate than $O(M^{-1/2})$. See \cite{lemieux2009monte}
for a review of such methods, \cite{chen2011} for the extension to
quasi-MCMC algorithms, and \cite{gerber:chopin:2015} for an entry on
sequential quasi-Monte Carlo.}

In contrast, when applying an \textit{approximate} method (using the
taxonomy in Section \ref{primer}), investigators make no claim to exactness,
other than citing the asymptotic (in $n$) accuracy of the Laplace
approximation-based methods (\citealp{tierney:kadane:1986}; 
\citealp{rue:martino:chopin:2009}%
), or the asymptotic validity of certain other approximations ({%
\citealp{fearnhead2018asymptotics}%
; \citealp{FMRR2016}; \citealp{frazier2019bayesian}; %
\citealp{zhang2017convergence}}). That is, for finite $n$ at least, such
methods are only ever acknowledged as providing an approximation to (\ref%
{gen_expect}), with that approximation perhaps claimed to be as accurate as
possible, given the relevant choice variables that characterize the method;
but no more than that.

So what benefits do such techniques offer, in return for sacrificing the
goal of exact inference? With reference to the methods discussed below: ABC
and BSL both completely obviate the need to evaluate $p(\mathbf{y}|%
\boldsymbol{\theta })$ and, in so doing, open up to Bayesian treatment a
swathe of empirical problems -- so-called \textit{doubly-intractable }%
problems\ -- in which neither $p(\boldsymbol{\theta }\mathbf{|y})$ \textit{%
nor }$p(\mathbf{y}|\boldsymbol{\theta })$ is available analytically;
problems that would otherwise not be amenable to Bayesian analysis. In
computing (\ref{gen_expect}), both methods replace the posterior in the
integrand, $p(\boldsymbol{\theta }\mathbf{|y})$, with an approximate
posterior based solely on \textit{simulation }from $p(\mathbf{y}|\boldsymbol{%
\theta })$ (and $p(\boldsymbol{\theta })$). A simulation-based estimate of (%
\ref{gen_expect}), $\overline{g(\boldsymbol{\theta }\mathbf{)}}%
=(1/M)\sum\nolimits_{i=1}^{M}g(\boldsymbol{\theta }^{(i)}\mathbf{)}$, is
then produced using draws, $\boldsymbol{\theta }^{(i)}$, from this
approximate posterior. In contrast, VB and INLA both require evaluation of $%
p(\mathbf{y}|\boldsymbol{\theta })$, but reap computational benefits in
certain types of problems (in particular those of high-dimension and/or
based on huge data sets) by replacing -- at least in part -- \textit{%
simulation} with (in some cases closed-form) \textit{optimization}. In the
case of VB, the posterior $p(\boldsymbol{\theta }\mathbf{|y})$ used to
define (\ref{gen_expect}) is replaced by an approximation produced via
calculus of variations. Depending on the nature of the problem, including
the variational family from which the optimal approximation is produced, the
integral is computed in either closed form or via a simulation step. With
INLA, the approximation of $p(\boldsymbol{\theta }\mathbf{|y})$ is chosen in
such a way that (\ref{gen_expect}) can be computed with a combination of
optimization and low-dimensional deterministic integration steps.

\subsection{Simulation-based approaches}

\subsubsection{Approximate Bayesian computation (ABC)\label{abc}}

From its initial beginnings as a practical approach for inference in
population genetics models with{\ computationally expensive likelihoods\ }(%
\citealp{tavare:balding:griffith:donnelly:1997}; %
\citealp{pritchard:seielstad:perez:feldman:1999}), ABC has grown in
popularity and is now commonly applied in numerous fields;{\ its broad
applicability highlighted by the more than }18,000 citations{\ garnered on
Google Scholar since 2000. }As such, not only do several reviews of the area
exist (e.g. \citealp{marin:pudlo:robert:ryder:2011}; %
\citealp{sisson2011likelihood}; {\citealp{lintusaari2017fundamentals}; %
\citealp{Beaumont2019}}), but the technique has recently reached `handbook
status', with the publication of \cite{sisson2018handbook}; and it is to
those resources that we refer the reader for extensive details on the
method, application and theory of ABC. We provide only the essence of the
approach here, including its connection to other computational methods.

The aim of ABC is to approximate $p(\boldsymbol{\theta }\mathbf{|y})$ in
cases where, despite the complexity of the problem preventing the \textit{%
evaluation} of $p(\mathbf{y|}\boldsymbol{\theta })$, $p(\mathbf{y|}%
\boldsymbol{\theta })$ (and $p(\boldsymbol{\theta })$) can still be \textit{%
simulated}. The simplest (accept/reject) form of the algorithm is {given in
Algorithm \ref{ABCalg1}}, {where} $d\{\cdot ,\cdot ,\}$ {denotes} a generic
metric and $\varepsilon >0$ a pre-specified or post-processing tolerance
parameter: 
\begin{algorithm}
	\caption{Vanilla Accept/Reject ABC Algorithm}
	\label{ABCalg1}
	\begin{algorithmic}
	\FOR{$i=1,\dots,M$}	
	\STATE Simulate $\boldsymbol{\theta }^{i}$, $i=1,2,...,M$, from $p(%
	\boldsymbol{\theta })$, and artificial data $\mathbf{z}^{i}$ from $p(%
	\boldsymbol{\cdot }|\boldsymbol{\theta }^{i})$;
	\STATE Accept $\boldsymbol{\theta }^{i}$ if $d\{\mathbf{z}%
	^{i},\mathbf{y}\}\leq \varepsilon $
	\ENDFOR 
	\end{algorithmic}	
\end{algorithm} 

An accepted $\boldsymbol{\theta }^{i}$ is a draw from the posterior: 
\begin{equation*}
p_{\varepsilon }(\boldsymbol{\theta }|\mathbf{y})=\frac{\int_{\mathcal{X}%
}p_{\varepsilon }(\boldsymbol{\theta },\mathbf{z}|\mathbf{y})\mathrm{d}%
\mathbf{z}}{\int_{\Theta }\int_{\mathcal{X}}p(\boldsymbol{\theta },\mathbf{z}%
|\mathbf{y})\mathrm{d}\mathbf{z}\mathrm{d}\boldsymbol{\theta }},\quad
p_{\varepsilon }(\boldsymbol{\theta },\mathbf{z}|\mathbf{y})=\mathbb{I}\left[
d\{\mathbf{y},\mathbf{z}\}\leq \varepsilon \right] p(\mathbf{z}|\boldsymbol{%
\theta })p(\boldsymbol{\theta }),
\end{equation*}%
where $\mathbb{I}\left[ \cdot \right] $ denotes the indicator function.\
Under regularity conditions, it can be shown that $\lim_{\varepsilon
\rightarrow 0}p_{\varepsilon }(\boldsymbol{\theta }|\mathbf{y})=p(%
\boldsymbol{\theta }|\mathbf{y})$. However, in practice the choice of $%
\varepsilon =0$ is infeasible since if $\mathbf{y}$ is continuous, the event 
$\mathbf{z}=\mathbf{y}$ has zero probability. More generally, for a fixed
computing budget, ensuring that $\alpha _{n}=\text{Pr}\{d\{\mathbf{z},%
\mathbf{y}\}\leq \varepsilon \}$ is non-negligible in practice for small $%
\varepsilon $ is infeasible as $n$ increases. Thus, unless we have very few
observations, or are working with discrete data, Algorithm \ref{ABCalg1}
cannot be implemented in anything but toy problems.

\paragraph{ABC using summary statistics \label{summ}\protect\bigskip \newline
}

Since comparing high-dimensional Euclidean vectors $\mathbf{z}$ and $\mathbf{%
y}$ is computationally infeasible, the vast majority of ABC applications
first degrade the datasets down to a vector of lower-dimensional statistics,
customarily called \emph{summary statistics}. Define ${\eta }:\mathcal{X}%
\rightarrow \mathcal{B}\subseteq \mathbb{R}^{k_{\eta }}$ as a summary
statistic mapping. In general then, Algorithm \ref{ABCalg1} is replaced
with: 
\begin{algorithm}
	\caption{Accept/Reject ABC Algorithm Based on Summary Statistics}
	\label{ABCalg2}
	\begin{algorithmic}
	\FOR{$i=1,\dots,M$}	
	\STATE Simulate $\boldsymbol{\theta }^{i}$, $i=1,2,...,M$, from $p(%
	\boldsymbol{\theta })$, and artificial data $\mathbf{z}^{i}$ from $p(%
	\boldsymbol{\cdot }|\boldsymbol{\theta }^{i})$;
	\STATE Accept $\boldsymbol{\theta }^i$ if $d\{\eta (\mathbf{z}^{i}),\eta (%
                 \mathbf{y})\}\leq \varepsilon $.
	\ENDFOR 
	\end{algorithmic}	
\end{algorithm}

In this more common formulation, ABC thus produces draws of $\boldsymbol{%
\theta }$ from a posterior that conditions not on the full data set $\mathbf{%
y}$, but on statistics $\eta (\mathbf{y})$ (with dimension less than $n$)
that summarize the key characteristics of $\mathbf{y.}$ Only if $\eta (%
\mathbf{y})$ are sufficient for conducting inference on $\boldsymbol{\theta }
$, and for $\varepsilon \rightarrow 0$, does ABC provide draws from the
exact posterior $p(\boldsymbol{\theta }\mathbf{|y})$. In practice, the
complexity of the models to which ABC is applied implies -- almost by
definition -- that {a {low-dimensional} set of sufficient statistics is
unavailable}, and the implementation of the method (in finite computing
time) requires a non-zero value for $\varepsilon $, and a given number of
draws, $M.$ Consequently, since $\varepsilon >0$, accepted draws from
Algorithm \ref{ABCalg2} can only be seen as draws from the posterior $%
p_{\varepsilon }(\boldsymbol{\theta }|\eta (\boldsymbol{y}))$, which is an
approximation to the `partial' posterior $p(\boldsymbol{\theta |}\eta (%
\boldsymbol{y}))$ that results from using a non-zero tolerance.\footnote{%
\citet{wilkinson:2013} argues that an equally valid interpretation of
Algorithm \ref{ABCalg2} is that it produces exact draws from a controlled
approximation to the target posterior, $p(\boldsymbol{\theta }|\eta (%
\boldsymbol{y}))$. This controlled approximation $p_{\varepsilon }(%
\boldsymbol{\theta }|\eta (\boldsymbol{y}))$ is actually expressible as a
convolution of the exact partial posterior with a kernel function that is
used to represent error in the summary statistics. This convolution can
itself be interpreted as an exact posterior associated with a randomised
version of $\eta (\boldsymbol{y})$.}

The accuracy of the posterior output by Algorithm \ref{ABCalg2} can be
understood via\textbf{\ }the decomposition: 
\begin{eqnarray}
\int_{\Theta }|p_{\varepsilon }(\boldsymbol{\theta }|\eta (\mathbf{y}))-p(%
\boldsymbol{\theta }|\mathbf{y})|\mathrm{d}\boldsymbol{\theta }
&=&\int_{\Theta }|p_{\varepsilon }(\boldsymbol{\theta }|\eta (\mathbf{y}))-p(%
\boldsymbol{\theta }|\eta (\mathbf{y}))+p(\boldsymbol{\theta }|\eta (\mathbf{%
y}))-p(\boldsymbol{\theta }|\mathbf{y})|\mathrm{d}\boldsymbol{\theta } 
\notag \\
&\leq &\int_{\Theta }|p_{\varepsilon }(\boldsymbol{\theta }|\eta (\mathbf{y}%
))-p(\boldsymbol{\theta }|\eta (\mathbf{y}))|\mathrm{d}\boldsymbol{\theta }%
+\int_{\Theta }|p(\boldsymbol{\theta }|\eta (\mathbf{y}))-p(\boldsymbol{%
\theta }|\mathbf{y})|\mathrm{d}\boldsymbol{\theta }.  \label{eq:deomp}
\end{eqnarray}%
The first term in \eqref{eq:deomp} captures the discrepancy between the
partial posterior we \textit{wish} to target, i.e., $p(\boldsymbol{\theta }%
|\eta (\mathbf{y}))$, and the posterior that is targeted by Algorithm \ref%
{ABCalg2}, i.e., $p_{\varepsilon }(\boldsymbol{\theta }|\eta (\mathbf{y}))$.
The second term measures the discrepancy that results from the use of
summary statistics $\eta (\mathbf{y})$ that are (most likely) insufficient
for the data $\mathbf{y}$.

{With regard to }the second term in \eqref{eq:deomp}, the difference{\ }is
characterized by the informativeness, or otherwise, of the chosen summaries.
Under regularity conditions that ensure both posteriors $p(\boldsymbol{%
\theta }|\eta (\mathbf{y}))$ and $p(\boldsymbol{\theta }|\mathbf{y})$ are
asymptotically Gaussian, as $n\rightarrow \infty $, and concentrate onto the
same value in $\Theta $, this difference is determined by the difference in
the posterior variances: in particular, for $n\rightarrow \infty $, 
\begin{equation*}
\int_{\Theta }|p(\boldsymbol{\theta }|\eta (\mathbf{y}))-p(\boldsymbol{%
\theta }|\mathbf{y})|^{2}\mathrm{d}\boldsymbol{\theta }\leq {\text{KL}\left[
p(\boldsymbol{\theta }|\mathbf{y}),p(\boldsymbol{\theta }|\eta (\mathbf{y}))%
\right] }\approx \frac{1}{2}\left[ \ln \frac{|\mathcal{I}_{\eta }|}{|%
\mathcal{I}|}-\text{dim}({\boldsymbol{\theta }})+\text{Tr}\left[ \mathcal{I}%
_{\eta }^{-1}\mathcal{I}\right] \right] ,
\end{equation*}%
where $\mathcal{I}$ (respectively, $\mathcal{I}_{\eta }$) denotes the Fisher
information matrix of the likelihood $p(\mathbf{y}|\boldsymbol{\theta })$
(respectively, $p(\eta (\mathbf{y})|\boldsymbol{\theta })$), $|\cdot |$ {%
denotes }the determinant, and $\text{Tr}(\cdot )$ the trace operator.
Clearly, the above is zero if and only if $\mathcal{I}_{\eta }=\mathcal{I}$,
i.e., if {and only if} the summaries are sufficient. More generally, the
above relationship demonstrates that the more informative {are} the
summaries, i.e., the closer $\eta (\mathbf{y})$ is to being sufficient, the
closer the partial posterior $p(\boldsymbol{\theta }|\eta (\mathbf{y}))$
will be to the exact posterior $p(\boldsymbol{\theta }|\mathbf{y})$. To this
end, some attention has been given to maximizing the information content of
the summaries in some sense (e.g. \citealp{joyce:marjoram:2008}; %
\citealp{blum:2010}; \citealp{fearnhead:prangle:2012}). This includes the
idea of defining $\eta (\mathbf{y})$ as (some function of) the maximum
likelihood estimator (MLE) of the parameter vector of an approximating
`auxiliary' model; thereby producing summaries that are -- via the
properties of the MLE -- close to being \textit{asymptotically} sufficient,
depending on the accuracy of the approximating model (%
\citealp{drovandi:pettitt:faddy:2011}; \citealp{drovandi2015bayesian}; %
\citealp{martin2019auxiliary}). This approach mimics, in the Bayesian
setting, the frequentist methods of indirect inference %
\citep{gourieroux:monfort:renault:1993} and efficient method of moments %
\citep{gallant1996moments} using, as it does, an approximating model to
produce feasible inference about an intractable true model. Whilst the price
paid for the approximation in the frequentist case is reduced sampling
efficiency, in the Bayesian case the cost is posterior inference that is
conditioned on insufficient summaries, and is partial inference as a
consequence.

Analyzing the first term in \eqref{eq:deomp}, we note that if $\varepsilon
\rightarrow 0$, then under reasonable assumptions, (such as, e.g., the
dominance condition $p_{\varepsilon }(\boldsymbol{\theta }|\eta (\mathbf{y}%
))\leq C<\infty $, for all $\boldsymbol{\theta }\in \Theta $ and all $%
\varepsilon \rightarrow 0$), this first term will converge to zero. However,
in practice, since Algorithm \ref{ABCalg2} generates $i.i.d.$ draws of $%
\boldsymbol{\theta }$ under the prior, many of the subsequent $\eta (\mathbf{%
z}^{i})$ values will be far away from $\eta (\mathbf{y})$; hence a large
value of $\varepsilon $ may be required to obtain a reasonable acceptance
rate for the algorithm. Consequently, obtaining draws from $p_{\varepsilon }(%
\boldsymbol{\theta }|\eta (\mathbf{y}))$ can be difficult as $\varepsilon $
becomes small. 
In the regime when $\varepsilon $ is large, the first term in equation %
\eqref{eq:deomp} can be large even if the second term in equation %
\eqref{eq:deomp} is small. To address {this issue}, several extensions of
the basic ABC algorithm have been proposed that seek to increase the mass of
simulated summaries $\eta (\mathbf{z})$ in the region of $\eta (\mathbf{y})$%
, with the hope being that these methods yield a more accurate approximation
to $p(\boldsymbol{\theta }|\eta (\mathbf{y}))$. These proposals broadly fall
into two classes, and are often used in conjunction: the first is the use of
post-processing corrections, the second is the use of proposals that `learn'
regions of $(\boldsymbol{\theta },\mathbf{z})$ where $\eta (\mathbf{z})$ is
closer to $\eta (\mathbf{y})$.\footnote{%
We make mention here of a further method that shares some features in common
with these two categories of method, namely Bayesian optimization for
likelihoood-free inference, or BOLFI (\citealp{gutmann2016bayesian}). BOLFI
uses Bayesian optimization to iteratively build a probabilistic model for
the relationship between $\boldsymbol{\theta }$ and the distance function $%
d\{\eta (\mathbf{y}),\eta (\mathbf{z})\}.$ The effect of this is to produce
draws that yield small values for $d\{\cdot ,\cdot \}$ and, hence, to reduce
the number of required model simulations. The principle is equally
applicable to the BSL technique to be discussed below.}

Broadly speaking, post-processing corrections adjust the accepted draws
obtained from an initial ABC algorithm according to a given model, the most
common being some form of regression model, in an attempt to increase the
accuracy of the posterior approximation at a fixed value of $\varepsilon $;
see, \citet{be02}, \citet{blum:2010}, \citet{blum:francois:2010} for
examples, and \cite{blum2019regression} for a review. Alternatively, methods
that `learn' proposal distributions can deliver more simulations in regions
of $(\boldsymbol{\theta },\mathbf{z})$ where $\eta (\mathbf{z})$ is close to 
$\eta (\mathbf{y})$. One such approach is to insert an MCMC step, and
associated proposal distribution, within Algorithm \ref{ABCalg2} in order to
more effectively explore the space, which yields an ABC-MCMC algorithm (%
\citealp{marjoram:etal:2003}). A more efficient exploration of the posterior
space for $(\boldsymbol{\theta },\mathbf{z})$ increases the likelihood that
we obtain draws of $\eta (\mathbf{z})$ closer to $\eta (\mathbf{y})$, and
subsequently ensures that, all else equal, ABC-MCMC can use {a smaller
tolerance than that }used in Algorithm 2, and thus obtain a more accurate
approximation to $p(\boldsymbol{\theta }|\eta (\mathbf{y}))$.

The downside of ABC-MCMC is that {unless the tolerance }$\varepsilon ${\ is
carefully tuned}, the resulting Markov {chain} can {mix poorly, thus leading
to} unreliable inference. Consequently, the use of approaches other than%
\textbf{\ }MCMC within Algorithm \ref{ABCalg2}, for instance based on a
decreasing sequence of tolerances, is commonplace. Indeed, arguably the most
popular current approach {to conducting }ABC inference is to insert
sequential, or `adaptive', proposals within Algorithm \ref{ABCalg2}, which
leads to ABC-SMC/ABC-population(P)MC algorithms; see %
\citet{sisson:fan:tanaka:2007} and \citet{beaumont:cornuet:marin:robert:2009}
for examples, and \citet{sissonfan2019} for a review. ABC-SMC learns
effective proposal distributions sequentially as part of the algorithm,
which, all else equal, can deliver better approximations to $p(\boldsymbol{%
\theta }|\eta (\mathbf{y}))$ than Algorithm \ref{ABCalg2}.{\ Importantly,
since }ABC-SMC is based on sequential importance sampling, the resulting {%
independent }posterior draws are free from the stickiness that can {arise }%
in ABC-MCMC. Furthermore, most common ABC-SMC algorithms sequentially learn
the tolerance $\varepsilon $ {also }so that explicit tuning of the tolerance
is not required.\footnote{{We refer to \cite{kousathanas2018guide} for a
review of software that enables many of the ABC algorithms discussed in this
section to be easily implemented.}}

{As a final point, we make note of the well-known curse of dimensionality to
which ABC is subject. At its simplest level, the estimation of }$%
p_{\varepsilon }(\boldsymbol{\theta }|\eta (\mathbf{y}))$ ({for any given }$%
\varepsilon ${) using the }$M$ {draws of Algorithm \ref{ABCalg2}, is
equivalent to nonparametric conditional density estimation. As such, the
accuracy of the estimate degrades as the dimension of }$\eta (\mathbf{y})$ {%
increases. Equivalently, a given level of accuracy requires a larger value
of }$M$ {and, hence, entails a higher computational burden, the larger is
the dimension of }$\eta (\mathbf{y})${. Whilst the modifications of ABC
noted above potentially reduce the computational burden associated with any
given }$\eta (\mathbf{y})${\ -- by either correcting draws post-simulation,
or producing more effective draws in the first place -- the issue of
dimension still obtains, and is simply intrinsic to the selection method
that underpins ABC. See \citet{blum:etal:2013} and \citet{nott2018high}} {%
for in-depth discussions, and also \citet{FMRR2016} for additional insights} 
{into the impact of the dimension of }$\boldsymbol{\theta }$ {on the
asymptotic behaviour of ABC.}

\paragraph{ABC using full data distances\protect\bigskip \newline
}

Recently, several researchers have begun to explore the use of ABC methods
that do not rely on summary statistics, but instead match empirical measures
calculated from the observed and simulated data using appropriate metrics.
In such cases, the accept/reject step in Algorithm \ref{ABCalg2} is simply
replaced with a discrepancy over the space of probability measures. More
formally, let $\hat{\mu}$ denote the empirical measure of the observed
sample and $\hat{\mu}_{\boldsymbol{\theta }}$ the empirical measure
calculated from the simulated sample $\mathbf{z}$. Then, for $\mathcal{D}(%
\hat{\mu}_{n},\hat{\mu}_{\boldsymbol{\theta }})$ denoting a generic
discrepancy that measures the difference between $\hat{\mu}_{n}$ and $\hat{%
\mu}_{\boldsymbol{\theta }}$, the distance between the summaries, $d\{\eta (%
\mathbf{y}),\eta (\mathbf{z})\}$, is replaced by $\mathcal{D}(\hat{\mu}_{n},%
\hat{\mu}_{\boldsymbol{\theta }})$.

Several choices of $\mathcal{D}(\cdot ,\cdot )$ have been proposed,
including the Wasserstein distance (\citealp{Bernton2019}), KL divergence (%
\citealp{jiang2018approximate}), minimum mean discrepancy (%
\citealp{park2016k2}), the energy distance (\citealp{nguyen2020approximate})
and the Cramer-von Mises distance (\citealp{frazier2020robust}). Recently, 
\cite{drovandi2021comparison} have undertaken an in-depth comparison of
these different methods for conducting inference, and compared\textbf{\ }the
results with a generic summary statistic-based ABC approach across several
examples. The authors' main findings are three-fold. First, the
distance-based approaches are {found to be }promising, {and to deliver}
reasonable inferences in many cases, {whilst obviating the need to seek} a
vector of informative summary statistics. {Secondly}, {and as a slight
qualification to the first finding,} {the authors find that distance-based
approaches must be combined with summary statistics to ensure identification
of }$\boldsymbol{\theta }$\textbf{\ }in certain classes of models. Lastly,
at least in their experiments, the best performing summary statistic-based
approach always {performs} at least as well as the best distance-based
approach, which suggests that if one can find informative summary statistics
they may outperform distance-based approaches in general.

\subsubsection{Bayesian synthetic likelihood (BSL)\label{bsl}}

Summary statistic-based ABC targets $p(\boldsymbol{\theta }\mathbf{|}\eta (%
\mathbf{y}))\propto p(\eta (\mathbf{y})\mathbf{|}\boldsymbol{\theta })p(%
\boldsymbol{\theta }),$ with $p(\boldsymbol{\theta }\mathbf{|}\eta (\mathbf{y%
}))$ itself, for insufficient $\eta (\mathbf{y})$, being an approximate
representation of $p(\boldsymbol{\theta }\mathbf{|y}).$ It is clear then
that, embedded within the {simplest accept/reject ABC algorithm}, based on a
tolerance $\varepsilon $, is a likelihood function of the form, 
\begin{equation}
p_{\varepsilon }(\eta (\mathbf{y})|\boldsymbol{\theta })=\int_{\mathcal{X}}p(%
\mathbf{z}|\boldsymbol{\theta })\mathbb{I}\left( d\{\eta (\mathbf{y}),\eta (%
\mathbf{z})\}\leq \varepsilon \right) d\mathbf{z}.  \label{abc_like}
\end{equation}%
For a given draw $\boldsymbol{\theta }^{i}$, and associated $\eta (\mathbf{z}%
^{i})$, (\ref{abc_like}) is approximated by its unbiased simulation
counterpart, $\widehat{p}_{\varepsilon }(\eta (\mathbf{y})|\boldsymbol{%
\theta }^{i})$ $=\mathbb{I}\left( d\{\eta (\mathbf{y}),\eta (\mathbf{z}%
^{i})\}\leq \varepsilon \right) ,$ which can implicitly be viewed as a
nonparametric estimator, based on a Uniform kernel, for the quantity of
interest $p_{\varepsilon }(\eta (\boldsymbol{\mathbf{y}})|\boldsymbol{\theta 
})$. Following \cite{andrieu:roberts:2009}, and as illustrated in detail by 
\cite{bornn2017use}, $\widehat{p}_{\varepsilon }(\eta (\mathbf{y})|%
\boldsymbol{\theta }^{i})$ can serve as a likelihood estimate within a form
of pseudo-marginal MCMC scheme (referred to as ABC-MCMC by the authors) for
sampling from $p_{\varepsilon }(\boldsymbol{\theta }\mathbf{|}\eta (\mathbf{y%
}))\propto p_{\varepsilon }(\eta (\mathbf{y})|\boldsymbol{\theta })p(%
\boldsymbol{\theta }),$ where in this context we take `pseudo-marginal MCMC'
to mean an MCMC scheme that replaces the intractable likelihood, $%
p_{\varepsilon }(\eta (\mathbf{y})|\boldsymbol{\theta })$, within a
Metropolis-Hastings ratio by an unbiased estimator, $\widehat{p}%
_{\varepsilon }(\eta (\mathbf{y})|\boldsymbol{\theta }^{i})$. (See also %
\citealp{marjoram:etal:2003}.) However, in contrast with other results in
the pseudo-marginal literature, \cite{bornn2017use} demonstrate that the
efficiency of the MCMC chain so produced is not necessarily improved by
using more than one draw of $\eta (\mathbf{z}^{i})$ for a given draw $%
\boldsymbol{\theta }^{i}.$

Bayesian synthetic likelihood (BSL)\textit{\ }\citep{price2018bayesian} also
targets a posterior for $\boldsymbol{\theta }$ that conditions on $\eta (%
\mathbf{y})$, and requires only\ simulation\textit{\ }from $p(\mathbf{y|}%
\boldsymbol{\theta })$\textit{\ }(not its evaluation) in so doing.\textit{\ }%
However, in contrast to the nonparametric likelihood estimate that is
implicit in ABC, BSL (building on \citealp{wood:2010}) overwhelmingly adopts
a Gaussian parametric approximation to $p(\eta (\mathbf{y})|\boldsymbol{%
\theta })$,%
\begin{equation}
p_{a}(\eta (\mathbf{y})|\boldsymbol{\theta })=\mathcal{N}\left[ \eta (%
\mathbf{y});\mu (\boldsymbol{\theta }),\Sigma (\boldsymbol{\theta })\right]
,\;\mu (\boldsymbol{\theta })=\mathbb{E}[\eta (\mathbf{y})],\;\Sigma (%
\boldsymbol{\theta })=\text{Var}\left[ \eta (\mathbf{y})\right] .
\label{bsl_likelihood}
\end{equation}%
Use of this parametric kernel leads to the \textit{ideal} BSL posterior, 
\begin{equation}
p_{a}(\boldsymbol{\theta }\mathbf{|}\eta (\mathbf{y}))\propto p_{a}(\eta (%
\mathbf{y})|\boldsymbol{\theta })p(\boldsymbol{\theta }),  \label{bsl_1}
\end{equation}%
where the subscript `$a$' highlights that (\ref{bsl_1}) is still an
approximation to $p(\boldsymbol{\theta }\mathbf{|}\eta (\mathbf{y}))$, due
to the Gaussian approximation, $p_{a}(\eta (\mathbf{y})|\boldsymbol{\theta }%
) $, of $p(\eta (\mathbf{y})\mathbf{|}\boldsymbol{\theta })$.

In general, however, the mean and variance-covariance matrix of $\eta (%
\mathbf{y})$ are unknown and must be estimated via simulation. Given $%
\mathbf{z}_{j}\sim i.i.d.$ $p(\boldsymbol{\cdot }|\boldsymbol{\theta })$, $%
j=1,\dots ,m$, we can estimate $\mu (\boldsymbol{\theta })$ and $\Sigma (%
\boldsymbol{\theta })$ in (\ref{bsl_likelihood}) via their empirical Monte
Carlo averages, $\mu _{m}(\boldsymbol{\theta })=\frac{1}{m}%
\sum_{j=1}^{m}\eta (\mathbf{z}_{j})$ and $\Sigma _{m}(\boldsymbol{\theta })=%
\frac{1}{m-1}\sum_{j=1}^{m}\mathcal{(}\eta (\mathbf{z}_{j})-\mu _{m}(%
\boldsymbol{\theta }))(\eta (\mathbf{z}_{j})-\mu _{m}(\boldsymbol{\theta }%
))^{\prime },$ and thereby define%
\begin{equation}
p_{a,m}(\eta (\mathbf{y})\mathbf{|}\boldsymbol{\theta })=\int_{\mathcal{X}}%
\mathcal{N}\left[ \eta (\mathbf{y});\mu _{m}(\boldsymbol{\theta }),\Sigma
_{m}(\boldsymbol{\theta })\right] \prod_{j=1}^{m}p(\eta (\mathbf{z}_{j})|%
\boldsymbol{\theta })d\mathbf{z}_{1}\dots d\mathbf{z}_{m},
\label{bsl_likelihood_2}
\end{equation}%
and the associated \textit{target} BSL posterior, 
\begin{equation}
p_{a,m}(\boldsymbol{\theta }\mathbf{|}\eta (\mathbf{y}))\propto p_{a,m}(\eta
(\mathbf{y})\mathbf{|}\boldsymbol{\theta })p(\boldsymbol{\theta }).
\label{bsl_2}
\end{equation}%
Note that, even for a single draw $\eta (\mathbf{z}_{j})$, $\mathbf{z}%
_{j}\sim p(\boldsymbol{\cdot }|\boldsymbol{\theta })$, we have that $%
\mathcal{N}\left[ \eta (\mathbf{y});\mu _{m}(\boldsymbol{\theta }),\Sigma
_{m}(\boldsymbol{\theta })\right] $ is an unbiased estimate of (\ref%
{bsl_likelihood_2}). Hence, with $p_{a,m}(\boldsymbol{\theta }\mathbf{|}\eta
(\mathbf{y}))$ then accessed via an MCMC algorithm, and with arguments in 
\cite{drovandi2015bayesian} used to show that $p_{a,m}\rightarrow p_{a}$ as $%
m\rightarrow \infty $, BSL can yield a form of pseudo-marginal\ MCMC method.
Pseudo-code for generic MCMC sampling of the BSL posterior in \eqref{bsl_2}
is given in Algorithm \ref{BSLalg1}. {We refer the interested reader to the
R language (\citealp{Rlang}) package BSL (\citealp{an2019bsl}), which can be
used to implement BSL{\ and its common variants}.}

\begin{algorithm}
	\caption{Vanilla BSL MCMC Algorithm}
	\label{BSLalg1}
	\begin{algorithmic}
		\FOR{$i=1,\dots,M$}	
		\STATE Draw $\bt^\ast\sim q(\bt|\bt^{i-1})$
		\STATE Produce $\mu_m(\bt)$ and $\Sigma_m(\bt)$ using $j=1,\dots,m$ independent model simulations at $\boldsymbol{\theta }^\ast$
		\STATE Compute the synthetic likelihood $L^\ast=\mathcal{N}\left[ \eta (\mathbf{y});\mu _{m}(\boldsymbol{\theta }^\ast),\Sigma
		_{m}(\boldsymbol{\theta }^\ast)\right]$ and $L^{i-1}$
		\STATE Compute the Metropolis-Hastings ratio:
		$$
		r=\frac{L^\ast\pi(\bt^\ast)q(\bt^{i-1}|\bt^\ast)}{L^{i-1}\pi(\bt^{i-1})q(\bt^\ast|\bt^{i-1})}
		$$
		\IF{$\mathcal{U}(0,1)<r$}
		\STATE Set $\bt^i=\bt^\ast$, $\mu_m(\bt^i)=\mu_m(\bt^\ast)$ and $\Sigma_m(\bt^i)=\Sigma_m(\bt^\ast)$	
		\ELSE
		\STATE Set $\bt^i=\bt^{i-1}$, $\mu_m(\bt^i)=\mu_m(\bt^{i-1})$ and $\Sigma_m(\bt^i)=\Sigma_m(\bt^{i-1})$			
		\ENDIF
		\ENDFOR 
	\end{algorithmic}	
\end{algorithm}

\subsubsection{ABC and BSL\label{vs}}

{Whilst (summary statistic-based) ABC and BSL target the same posterior, $p(%
\boldsymbol{\theta }|\eta (\mathbf{y}))$, both methods produce posteriors
that differ from this target,} and from one another{. Therefore, it is
helpful to characterize the difference between these posteriors in in terms
of their \textit{i)} large sample (in $n$) behaviour and \textit{ii)}}
computational efficiency. This then enables us to provide some guidelines as
to when, and why, one might use one method over the other. We consider 
\textit{i)} and\textit{\ ii)} in turn.

\textit{i) }As ABC has evolved into a common approach to inference,
attention has turned to its asymptotic validation. This work demonstrates
that, under certain conditions on $\eta (\mathbf{y})$, $\varepsilon $ and $M$%
, {as }$n\rightarrow \infty ${, the ABC posterior} ${p}_{\varepsilon }(%
\boldsymbol{\boldsymbol{\theta }}\mathbf{|}\eta (\mathbf{y})) $ targeted by
Algorithm \ref{ABCalg2}: concentrates onto the true vector $\boldsymbol{%
\boldsymbol{\theta }}_{0}$ (i.e.~is Bayesian consistent); satisfies a
Bernstein von Mises (BvM) theorem (i.e.~is asymptotically Gaussian) with
credible sets that have the correct level of frequentist asymptotic
coverage; and yields an ABC posterior mean with an asymptotically Gaussian
sampling distribution. (See \citealp{FMRR2016}, for this full suite of
results, and \citealp{LF2016b}, \citealp{LF2016a}, and %
\citealp{frazier2020model}, for related work.) Moreover, the conditions on $%
\eta (\mathbf{y})$ under which these results are valid are surprisingly
weak, requiring only the existence of at least a polynomial moment
(uniformly in the parameter space). In addition, the ABC posterior can be as
efficient as the maximum likelihood estimator based on the likelihood $%
p(\eta (\mathbf{y})|\boldsymbol{\theta })$.

The required conditions on the tolerance, $\varepsilon $, for these results
to be in evidence can be ordered in terms of the speed with which $%
\varepsilon \rightarrow 0$ as $n\rightarrow \infty $: stronger results, such
as a valid BvM, require faster rates of decay for $\varepsilon $ than weaker
results, such as posterior concentration. Such a taxonomy is important since
the chosen tolerance $\varepsilon $ largely determines the computational
effort required for ${p}_{\varepsilon }(\boldsymbol{\theta }|\eta (\mathbf{y}%
))$ to be an accurate estimate of $p(\boldsymbol{\theta }|\eta (\mathbf{y}))$%
. Broadly speaking, the smaller is\textbf{\ }$\varepsilon $, the smaller is%
\textbf{\ }$|{p}_{\varepsilon }(\boldsymbol{\theta }|\eta (\mathbf{y}))-p(%
\boldsymbol{\theta }|\eta (\mathbf{y}))|$. However, a smaller choice of $%
\varepsilon $ requires a larger number of simulations (i.e., a larger value
of $M$) and, hence, a greater computational effort. For instance, if we wish
for credible sets obtained by ${p}_{\varepsilon }(\boldsymbol{\theta }|\eta (%
\mathbf{y}))$ to be valid in the frequentist sense, $M$ is required to
diverge faster than $n^{\text{dim}(\boldsymbol{\eta })/2}$ (Corollary 1 in %
\citealp{FMRR2016}).

In contrast to ABC, BSL is based on the Gaussian approximation to the
likelihood $p(\eta (\mathbf{y})|\boldsymbol{\theta })$, and does not require
any choice of tolerance. However, in order for the BSL posterior $p_{a,m}(%
\boldsymbol{\theta }|\eta (\mathbf{y}))$ to be a reasonable approximation to 
$p(\boldsymbol{\theta }|\eta (\mathbf{y}))$, the Gaussian approximation must
be reasonable. More specifically, the summaries $\eta (\mathbf{y})$ and $%
\eta (\mathbf{z})$ themselves must satisfy a CLT (uniformly in the case of
the latter) (see \citealp{frazier2019bayesian}, for details), and the
variance of the summaries must be consistently estimated by $\Sigma _{m}(%
\boldsymbol{\theta })$ for some value of $\boldsymbol{\theta }$, as $m$ (the
number of data sets drawn for a given draw of $\boldsymbol{\theta }$)
increases. If, moreover, we wish $p_{a,m}(\boldsymbol{\theta }|\eta (\mathbf{%
y}))$ to deliver asymptotically correct frequentist coverage, additional
conditions on the summaries and $m$ are required. In particular, \cite%
{frazier2019approximate} demonstrate that if the summaries exhibit an
exponential moment, then correct uncertainty quantification is achieved so
long as $m/\log (n)$\textbf{$\rightarrow \infty $}. Under the restrictions
delineated above for $\eta (\mathbf{y})$, $\varepsilon $, $M$ and $m$, the
results of \cite{FMRR2016} and \cite{frazier2019bayesian} can then be used
to deduce that the ABC and BSL posteriors are asymptotically equivalent, in
the sense that $\int |p_{a,m}(\boldsymbol{\theta }|\eta (\mathbf{y}))-{p}%
_{\varepsilon }(\boldsymbol{\theta }|\eta (\mathbf{y}))|d\boldsymbol{\theta }%
\overset{p}{\rightarrow }0$ as $n\rightarrow \infty .$ That is, in large
samples, and under regularity, we could expect the results obtained by both
methods to be comparable. However, the above discussion makes plain that BSL
requires much stronger conditions on the summaries than does ABC to produce
equivalent asymptotic behaviour. Hence, in the case of summaries that have
thick tails, non-Gaussian features, or non-standard rates of convergence,
ABC would seem to be the better choice.

\textit{ii) }The above asymptotic comparison between ABC and BSL {abstracts
from the actual sampling required} to obtain draws from the posterior
targets; that is, the large sample behavior discussed above is divorced from
the actual practice of obtaining draws from the different posteriors, and
thus {ignores} the computational efficiency of the two approaches. Once
computational efficiency, is taken into account, the comparison between the
two methods becomes more nuanced. \cite{frazier2019bayesian} use theoretical
arguments to compare the computational efficiency of BSL and accept/reject
ABC, and demonstrate that BSL does not pay the same penalty for summary
statistic dimension as does ABC. In particular, the BSL acceptance
probability is asymptotically non-vanishing, and does not depend on the
dimension of the summaries, neither of which is true for accept/reject ABC,
even under an optimal choice for $M$. Given this, when the summaries are
approximately Gaussian, BSL is likely to be more computationally efficient
than standard ABC.\footnote{%
BSL can often be implemented using the random walk MH algorithm, and often
with minimal tuning required in practice (\citealp{price2018bayesian}). See
also \cite{frazier2019robust} for a slice sampling approach to sampling the
BSL posterior.}

\subsubsection{Illustrative example: ABC and BSL\label{ill_opt}}

We complete this section {on simulation-based approximate methods }with a {%
brief }discussion of an empirical example from \cite{drovandi2021comparison} 
{in which both ABC and BSL methods are applied.} We have selected this
particular example as our illustration because it is has two features that
are common to many empirical applications of {ABC and BSL}: \textit{1)} The
model does not enable a likelihood function to be computed {analytically},
but the model \textit{can} be simulated; \textit{2)} Despite the complexity
of the model, the number of parameters of interest is small; hence a
reasonably {small number of }summary statistics are able to be selected. The
illustration also includes a comparison of summary-statistic {based ABC with
ABC based on full distances.} {We present certain graphical output (Figure 5
in their original paper) as Figure \ref{fig1} below.}

The empirical {problem} is one of conducting inference on the large-valued
imperfections (or `inclusions') in steel that can arise during the
production process; or in general parlance, one of conducting inference for 
\textit{stereological extremes. }We refer to \cite{bordot2007}{\ }for all
details of the physical and statistical problems. Suffice to say, for the
illustrative purpose here, that a realistic model for explaining such
extreme {inclusions}, namely an ellipsoid family for {inclusion} shapes,
does not have {an available} likelihood function, but can be inexpensively
simulated. Moreover, the particular model analyzed in \cite%
{drovandi2021comparison} is described by only three parameters: the rate
parameter ($\lambda $) of a homogenous Poisson process describing the {%
random number} of {inclusions per volume of steel}, and the scale ($\sigma $%
) and shape ($\xi $) parameters of a generalized Pareto distribution related
to the size of the {inclusions}.

\cite{drovandi2021comparison} consider ABC based on two different sets of
summary statistics. The first choice is based on a similar set of four
statistics to that used in \cite{bordot2007} ({`ABC 4stats'} in Figure \ref%
{fig1}), while {the second set is based} on the nine-dimensional{\ }score
vector of an auxiliary Gaussian mixture model with three components ({`ABC
Summ'} in Figure \ref{fig1}); BSL-based inference is {based on this second
set of summaries only} (`BSL' in Figure \ref{fig1}). When applying the
distance-based ABC approaches, {\cite{drovandi2021comparison}} {note that
the inclusion size, a continuous variable, and the number of inclusions, a
discrete variable, both carry identifying information about the unknown
parameters.} To this end, the authors combine two distance functions, one
for the number of inclusions, and one for the inclusion sizes. For the
inclusion sizes, the authors use a range of distance functions {including}
Cramer-von mises (CvM), Wasserstein (Wass), {maximum mean discovery (}MMD),
and {the} simulation-based kernel density approach of \cite%
{turner2014generalized} (KDE). Each distance is then combined with the
absolute difference between the observed number of inclusions and the
simulated number of inclusions from the model.

\begin{figure}[h]
\centering
\subfigure[full data
distances]{\includegraphics[scale=0.7]{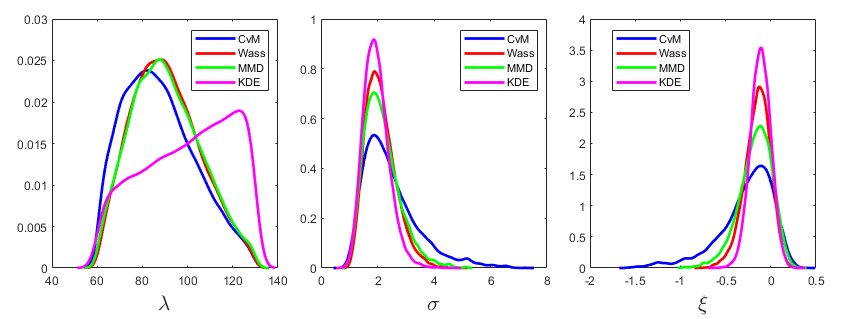}\label{figsub:results_stereo_real_dist}}
\subfigure[summary
statistics]{\includegraphics[scale=0.7]{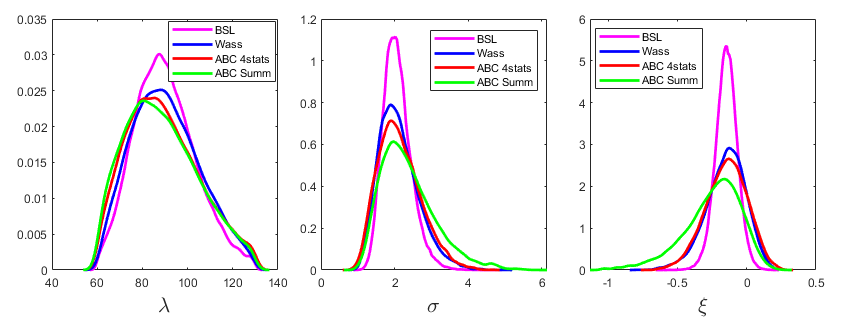}\label{figsub:results_stereo_real_summ}}
\caption{Figure 5 from \protect\cite{drovandi2021comparison}, with caption:
\textquotedblleft Comparison of estimates of the univariate ABC posterior
distributions for the stereological extremes example based on real data.
Shown are (a) comparisons with distance functions involving the full data
and (b) comparisons with summary statistic-based approaches."}
\label{fig1}
\end{figure}

Some key messages to be taken from Figure \ref{fig1} are as follows:\medskip

\begin{enumerate}
\item The ABC posteriors based on different summary {statistics} and
distance functions produce different posteriors! {More specifically, and as
is reasonably typical, }the posteriors {for any given parameter }are
generally centred {at} {similar points in the parameter space}, but have
varying {degrees of dispersion}. {Of} the {posteriors }based on summary
statistics, {plotted in the bottom row of the figure, }ABC based on {the
nine summaries derived from }the Gaussian mixture model {(`ABC summ') }has
the largest dispersion {in each case}. This reflects the {curse of
dimensionality in the dimension of the summary statistics to which ABC is
subject}, {as discussed in Section \ref{summ}.}

\item {Following on from the above point, and with reference to Point }%
\textit{ii)}\textbf{\ }in Section \ref{vs}, the BSL posterior based on the
Gaussian mixture model summaries is notably less dispersed than the
corresponding `ABC summ'. This difference can be attributed to the
approximate Gaussianity of the summary statistics in this example, which
results in a BSL posterior that is less sensitive to the dimension of the
summaries than ABC. Consequently, given the same computing budget for both
methods, we would expect that BSL would produce more efficient posteriors
since its acceptance rate does not decline as sharply as that of ABC when
the dimension of the summaries is moderate or large.

\item {With reference to the plots in the top row of Figure \ref{fig1}, not}
all distance functions produce reasonable posteriors. Like summaries,
different distances capture different features of the data. Moreover, as
mentioned above, the use of {a single }distance alone may not be able to
identify all models parameters in all circumstances. {Therefore, careful
preliminary analysis} should be {undertaken }when using distance-based ABC.

\item Lastly, the {least dispersed} summary statistic method ({i.e. }BSL)
has less {dispersion} than the best distance-based ABC approach ({`Wass'} in
this case). \cite{drovandi2021comparison} find similar behavior in all the
examples considered in their analysis, which suggests that, while
distance-based ABC approaches are useful as they obviate the crucial choice
of which summaries to select, they may not perform as well as methods based
on informative summary statistics, at least in cases where a feasible
informative and low-dimensional summary exists.
\end{enumerate}

\subsection{Optimization approaches}

\subsubsection{Variational Bayes (VB)\label{vb}}

The two approximate methods discussed thus far, ABC and BSL, target an
approximation of the posterior that is (in a standard application of the
methods) conditioned on a vector of low-dimensional summary statistics. As
such, and most particularly when $\eta (\mathbf{y})$ is not sufficient for $%
\boldsymbol{\theta }$, these methods do not directly target the exact
posterior $p(\boldsymbol{\theta }|\mathbf{y})$, nor any expectation, (\ref%
{gen_expect}), defined with respect to it. In contrast, VB methods are a
general class of algorithms that produce an approximation to the posterior $%
p(\boldsymbol{\theta }|\mathbf{y})$ -- and hence (\ref{gen_expect}) -- 
\textit{directly}, by replacing simulation with optimization.

The idea of VB is to search for the best approximation to the posterior $p(%
\boldsymbol{\theta }|\mathbf{y})$ over a class of densities $\mathcal{Q}$,
referred to as the variational family, and where $q(\boldsymbol{\theta })$
indexes elements in $\mathcal{Q}$. The most common approach to VB is to find
the best approximation to the exact posterior, in the class $\mathcal{Q}$,
by minimizing the KL divergence between $q(\boldsymbol{\theta })$ and the
posterior $p(\boldsymbol{\theta }|\mathbf{y})$, which defines such a density
as the solution to the following optimal optimization problem, 
\begin{equation}
q^{\ast }(\boldsymbol{\theta }):=\argmin_{q(\boldsymbol{\theta })\in 
\mathcal{Q}}\text{KL}\left[ q(\boldsymbol{\theta })|p(\boldsymbol{\theta }|%
\mathbf{y})\right] ,  \label{opt_1}
\end{equation}%
where 
\begin{equation}
\text{KL}\left[ q(\boldsymbol{\theta })|p(\boldsymbol{\theta }|\mathbf{y})%
\right] =\int \log (q(\boldsymbol{\theta }))q(\boldsymbol{\theta })d%
\boldsymbol{\theta }-\int \log (p(\boldsymbol{\theta }|\mathbf{y}))q(%
\boldsymbol{\theta })d\boldsymbol{\theta }\equiv \mathbb{E}_{q}[\log (q(%
\boldsymbol{\theta }))]-\mathbb{E}_{q}[\log (p(\boldsymbol{\theta },\mathbf{y%
}))]+\log (p(\mathbf{y}))  \label{KL}
\end{equation}%
{and }$p(\boldsymbol{\theta },\mathbf{y})=p(\mathbf{y}|\boldsymbol{\theta }%
)p(\boldsymbol{\theta }).$ Of course, the normalizing constant $p(\mathbf{y}%
) $ is, in all but most simple problems (for which VB would not be
required!), unknown; and the quantity in (\ref{KL}) inaccessible as a
result. Rather, the approach adopted is to define the so-called evidence
lower bound (ELBO),%
\begin{equation}
\text{ELBO}[q(\boldsymbol{\theta })]:=\mathbb{E}_{q}[\log (p(\boldsymbol{%
\theta },\mathbf{y}))]-\mathbb{E}_{q}[\log (q(\boldsymbol{\theta }))],
\label{elbo}
\end{equation}%
where KL$\left[ q(\boldsymbol{\theta })|p(\boldsymbol{\theta }|\mathbf{y})%
\right] $ is equivalent to $-$ELBO$\left[ q(\boldsymbol{\theta })\right] $
up to the unknown constant, $\log (p(\mathbf{y}))$,\ with the latter not
dependent on $q(\boldsymbol{\theta })$. Hence, we can obtain the variational
density by solving an optimization problem that is equivalent to that in (%
\ref{opt_1}): 
\begin{equation}
q^{\ast }(\boldsymbol{\theta }):=\argmax_{q(\boldsymbol{\theta })\in 
\mathcal{Q}}\text{ELBO}[q(\boldsymbol{\theta })].  \label{vb_1}
\end{equation}%
In practice,\textbf{\ }$q(\boldsymbol{\theta })$ is either explicitly or
implicitly parameterized by a vector of `variational parameters', $%
\boldsymbol{\lambda }$, and optimization occurs with respect to\textbf{\ }$%
\boldsymbol{\lambda }.$

The beauty of VB is that, for \textit{certain} problems, including certain
choices of the class $\mathcal{Q}$, the optimization problem in (\ref{vb_1})
can either yield a closed-form solution, or be solved relatively quickly
with various numerical algorithms; (see \citealp{ormerod2010explaining}, %
\citealp{blei2017variational}, and \citealp{zhang2018advances}, for
reviews). Most importantly, given that -- by design -- the variational
family is defined in terms of standard forms of distributions, replacement
of $p(\boldsymbol{\theta }|\mathbf{y})$\ by $q^{\ast }(\boldsymbol{\theta })$
in (\ref{gen_expect}) yields an expectation that is either available in
closed form, or amenable to a relatively simple simulation-based solution. {%
Moreover, the link between (\ref{KL}) and (\ref{elbo}) makes it clear that
maximizing (\ref{elbo}) to yield }$q^{\ast }(\boldsymbol{\theta })${\
produces, as a by-product, a lower bound on the logarithm of the `evidence',
or marginal likelihood, }$p(\mathbf{y}).${\ Hence, ELBO}$\left[ q^{\ast }(%
\boldsymbol{\theta })\right] ${\ serves as an estimate of the quantity that
underpins model choice.}

{The production of }$q^{\ast }(\boldsymbol{\theta })$, {and the associated
estimate of an approximation of (\ref{gen_expect}) as based on }$q^{\ast }(%
\boldsymbol{\theta })$, {is typically \textit{much} faster (often orders of
magnitude so) than producing an estimate of (\ref{gen_expect})} {via exact
simulation of }$p(\boldsymbol{\theta }|\mathbf{y}).$ {This is of particular
import when both }$\boldsymbol{\theta }$, {and possibly }$\mathbf{y}$ {also},%
{\ are high-dimensional. In such cases, the computational cost of simulating
from }$p(\boldsymbol{\theta }|\mathbf{y})${, via MCMC for example,} {may
simply be prohibitive, given the need to both explore a high-dimensional and
complex parameter space and -- at each point in that search -- evaluate }$p(%
\mathbf{y|}\boldsymbol{\theta })$ {at }$\mathbf{y}.${\ In contrast,} the {%
variational }family $\mathcal{Q}$, {and the optimization algorithm, }can be
chosen in such a way that a {VB approximation of} $p(\boldsymbol{\theta }|%
\mathbf{y})$ {can be produced within an acceptable timeframe,} even when the
dimension of $\boldsymbol{\theta }$ is in the thousands, or the tens of
thousands (\citealp{braun2010variational}; \citealp{kabisa2016online}; %
\citealp{wand2017fast}; \citealp{koop2018variational}). {The ability of VB
to scale to large models and datasets also makes the method particularly
suitable for exploring multiple models quickly, perhaps as a preliminary
step to a more targeted analysis (\citealp{blei2017variational}).}

We now give specific algorithmic details for two foundational VB algorithms:
coordinate ascent variational inference (CAVI) (see %
\citealp{bishop2006pattern}, Chapter 10, for discussion) and stochastic
variational inference (SVI) (\citealp{hoffman13}), both of which seek to
solve the optimization problem in (\ref{vb_1}), for given specifications of $%
p(\boldsymbol{\theta },\mathbf{y})$ and choices of $Q$. These algorithms
suit the intended purpose of this review as they both played a prominent
role in the initial development of the VB literature, and allow us to
discuss some of the mechanics of VB without getting needlessly bogged down
in the details. {For a review of more recent developments in VB, including
details of implementation, we refer to \cite{zhang2018advances}}.\footnote{%
We note that, unlike approximate Bayesian methods based on simulation, the
diverse, and complex, nature of the problems to which VB methods are applied
make it somewhat less well-suited to generating well-behaved, and reliable,
software products that can be used to implement {the methods} across a wide
range of problems. That being said, the automatic differentiation
variational inference (ADVI) approach of \cite{kucukelbir2017automatic} can
be implemented in many different problems, and is the default method for
variational inference in the popular probabilistic programming language STAN
(\citealp{carpenter2017stan}).}

The CAVI algorithm is derived for the `mean-field' variational family, where
the elements of $\boldsymbol{\theta }=(\theta _{1},\theta _{2},..,\theta
_{p})^{\prime }$ are specified as mutually independent, with joint density $%
q(\boldsymbol{\theta })=\prod\limits_{j=1}^{p}q_{j}(\theta _{j})$ denoting a
generic element of $Q$. The CAVI algorithm makes use of the fact that, under
the mean-field family, the density $q_{j}^{\ast }(\theta _{j})$, the
solution to \eqref{vb_1} for the $j$-th element of $\boldsymbol{\theta }$,
has the closed form $q_{j}^{\ast }(\theta _{j})\propto \exp ({\mathbb{E}%
_{-j}[log(p(}\theta _{j}|\boldsymbol{\theta }_{-j}{,\mathbf{\ y})]})$ -- {%
where }$\mathbb{E}_{-j}$ {denotes the expectation with respect to the
variational density over }$\boldsymbol{\theta }_{-j}$, $\prod\limits_{l\neq
j}q_{l}(\theta _{l})$ -- {which can be derived from (\ref{elbo}) by
exploiting the independence of the }$\theta _{j}$ {under the mean-field
family (see \citealp{blei2017variational}, p.10). However, this solution is
not explicit since $q_{j}^{\ast }(\theta _{j})$ depends on expectations
computed with respect to the other factors $q_{-j}(\boldsymbol{\theta }%
_{-j}) $. Hence, given an initial solution, CAVI cycles through $q_{j}^{\ast
}(\theta _{j})$, $j=1,\dots ,\text{dim}(\boldsymbol{\theta })$, updating
each factor in turn. {The fact that we are able to calculate $\mathbb{E}%
_{-j}[\log p(\theta _{j}|\boldsymbol{\theta }_{-j},\mathbf{y})]$ in closed
form ensures, in turn, that the algorithm provides a very speedy solution to
(\ref{vb_1}).}} In Algorithm \ref{CAVI}, we provide pseudo-code for
implementing CAVI, deferring to \cite{blei2017variational} for further
details.

{In contrast to CAVI, SVI is applicable to a broader range of scenarios for
both }$p(\boldsymbol{\theta },\mathbf{y})$ {and} $\mathcal{Q}$ ({see} {%
\citealp{hoffman13}, Section 5, on this point}){.} {In addition, it scales
better to very large data sets as, unlike CAVI, it does not require the full
vector }$\mathbf{y}$ {to be processed on each iteration. In Algorithm \ref%
{SVI}, we provide pseudo-code for implementing SVI for the case of a
`conditionally conjugate model' and a mean-field variational family, in
which we now exploit the breakdown of }$\boldsymbol{\theta }$ {into a vector
of global parameters, }$\boldsymbol{\phi }$, and {an }$n${-dimensional
vector of local parameters,} $\mathbf{x}${\ (see Section \ref{intract}).} {%
Referring to \citeauthor{blei2017variational} and \citeauthor{hoffman13} for
further details (and noting the differing notation), we assume the following
structure for the joint distribution:}%
\begin{equation}
p(\boldsymbol{\phi },\mathbf{x},\mathbf{y})=p(\boldsymbol{\phi }|\boldsymbol{%
\alpha })\prod\limits_{i=1}^{n}p(x_{i},y_{i}|\boldsymbol{\phi }),
\label{ccm}
\end{equation}%
{where }$p(x_{i},y_{i}|\boldsymbol{\phi })$ is a member of the linear
exponential family, and\textbf{\ }$p(\boldsymbol{\phi }|\boldsymbol{\alpha }%
) $ {is the appropriate natural conjugate prior, with hyperparameter vector, 
}$\boldsymbol{\alpha }$.{\ In the algorithm, }$\psi _{i}$ {denotes the
variational parameter for each local parameter, }$x_{i}$, $\boldsymbol{%
\lambda }$ {the vector of variational parameters associated with} $%
\boldsymbol{\phi }$, {and }$\eta (\cdot )$ {and} $t(\cdot )$ are specific
functional forms that define the member of the exponential family underlying
the specification in (\ref{ccm}) (see \citeauthor{blei2017variational}). The
key implication of the assumed exchangeable structure in \eqref{ccm} is that
this structure permits the use of stochastic optimization routines to search
the variational parameters $\boldsymbol{\lambda }$ that deliver the best
approximation in the class $\mathcal{Q}$. That is, in contrast with
Algorithm \ref{CAVI}, the full vector $\mathbf{y}$ need not be processed at
each iteration. Instead, a {single observation,} $y_{i}$, or batches of $%
y_{i}$, can be randomly selected and used to optimize the ELBO over both the
local and global variational parameters. This simplification allows the
algorithm to successfully scale to problems in which $\mathbf{y}$ is truly
massive, at the cost of assuming the exchangeable structure in \eqref{ccm}.

\begin{algorithm}
	\caption{Vanilla CAVI Algorithm}
	\label{CAVI}
	\begin{algorithmic}
		\STATE Input: A joint probability distribution, $p(\boldsymbol{\theta},\mathbf{y})=p(\mathbf{y}|\boldsymbol{\theta})p(\boldsymbol{\theta})$  
		\STATE Output: A variational density from the mean-field class, $q^{\ast }(%
		\boldsymbol{\theta })=\prod\limits_{j=1}^{p}q^\ast_{j}(\theta_{j} )$ 
		\STATE Initialize: Variational factors $q_{j}(\theta_{j} ), j=1,\dots,p$ 
		\WHILE {the ELBO in (\ref{elbo}) has not converged}
		\FOR {$j=1,\dots,p$}
		\STATE Set $q^\ast_{j}(\theta_{j} )\propto exp({\mathbb{E}_{-j}[log(p(\theta_{j}|\boldsymbol{\theta }_{-j},\mathbf{y}))]})$	
		\ENDFOR 
		\ENDWHILE
		\STATE Return: $q^{\ast }(\boldsymbol{\theta })$
		
	\end{algorithmic}	
\end{algorithm}

\begin{algorithm}
	\caption{SVI Algorithm for a Conditionally Conjugate Model}
	\label{SVI}
	\begin{algorithmic}
		\STATE Input: A joint probability distribution, $p(\boldsymbol{\phi },\mathbf{x},\mathbf{y})$ of the form specified in (\ref{ccm})
		\STATE Output: A variational density for the global parameters $q_{\boldsymbol{\lambda }}(\boldsymbol{\phi })$  
		\STATE Initialize: A variational parameter vector, $\boldsymbol{\lambda }^{(0)}$ 
		\STATE Set: The step-size schedule $\rho^{(t)}$
		\WHILE {TRUE}
		\STATE Choose a data point, $y_{i}$, uniformly at random, $i \sim  Unif(1,...,n)$
		\STATE Optimize its local variational parameter, $\psi _{i} =\mathbb{E}_{\boldsymbol{\lambda }^{(t-1)}}[\eta (\boldsymbol{\phi }%
		,y_{i})]$
		\STATE Compute the intermediate global variational parameter vector as though $y_{i}$ is replicated $n$ times, $\widehat{\boldsymbol{\lambda }} =\boldsymbol{\alpha} +n[\mathbb{E}_{\psi _{i}}[t(x_{i},y_{i})]',1]'$
		\STATE Update the global variational parameter vector, $\boldsymbol{\lambda }^{(t)}=(1-\rho ^{(t)})\boldsymbol{\lambda }%
		^{(t-1)}+\rho^{(t)}\widehat{\boldsymbol{\lambda }}$
	
		\ENDWHILE
		\STATE Return: $q_{\boldsymbol{\lambda }}(\boldsymbol{\phi })$
		
	\end{algorithmic}	
\end{algorithm}

{Recently,} several authors have analyzed the asymptotic properties of VB
methods; see, for example, \cite{wangblei2019b,wangblei2019a}, and \cite%
{zhang2017convergence}. The most complete treatment can be found in %
\citeauthor{zhang2017convergence}, wherein the authors demonstrate that the
rate at which the VB posterior concentrates is bounded above by the
following two components: \textit{i)} the concentration rate of the exact
posterior, and \textit{ii)} the approximation error incurred by the chosen
variational family. This novel decomposition highlights the fundamental
importance of the variational family that is used to approximate the
posterior, something that is not present in other results on the asymptotic
behavior of VB. Interestingly, while \citeauthor{zhang2017convergence}
deliver a convenient upper bound in a general context, they also demonstrate
that in specific examples, such as Gaussian sequence models and sparse
linear regression models, the VB posterior can display concentration rates
that are actually faster than those obtained by the exact posterior, owing
to the fact that VB performs a type of `internal regularization' as a
consequence of the algorithm's optimization step. As a final point, we note
that \cite{yao18} and \cite{huggins2019validated} propose methods for
validating the accuracy of VB posterior approximations using alternative
(nonasymptotic) principles.\footnote{%
See also \cite{yu2019assessment} (and earlier references therein) for
practical validation approaches that are relevant to approximate posteriors
in general.}

\subsubsection{Integrated nested Laplace approximation (INLA)\label{inla}}

In 1774, Pierre Simon Laplace published one of his many remarkable papers, 
\textit{`M\'{e}moire sur la probabilit\'{e} des causes par les \'{e}v\'{e}%
nemens'}, in which he produced the first asymptotic (in $n$) approximation
to a posterior probability.\footnote{%
See \cite{stigler:1975}, \cite{stigler:1986}, {\cite{stigler:Laplace1774}}
and \cite{fienberg:2006} {for various details about Laplace's role in the
development of} `inverse probability', or Bayesian inference as it is now
known.} In brief, and using a scalar $\theta $ for the purpose of
illustration, his original method can be explained as follows. Begin by
expressing an arbitrary posterior probability as%
\begin{equation}
\mathbb{P}(a<\theta <b|\mathbf{y})=\int\limits_{a}^{b}p(\theta \mathbf{|y}%
)d\theta =\int\limits_{a}^{b}\exp \left\{ nf(\theta )\right\} d\theta ,
\label{la_prob}
\end{equation}%
where $f(\theta )=\log \left[ p(\theta \mathbf{|y})\right] /n$, and assume
appropriate regularity for $p(\mathbf{y}|\theta )$ and $p(\theta ).$ What is
now referred to as the \textit{Laplace asymptotic approximation} involves
first taking a second-order Taylor series approximation of $f(\theta )$
around its mode, $\widehat{\theta }$: $f(\theta )\approx f(\widehat{\theta }%
)+\frac{1}{2}f^{^{\prime \prime }}(\widehat{\theta })(\theta -\widehat{%
\theta })^{2}$, where $f^{\prime }(\widehat{\theta })=0$ by construction.
Defining $\sigma ^{2}=-[nf^{^{\prime \prime }}(\widehat{\theta })]^{-1}$, {%
and }substituting {the expansion }into (\ref{la_prob}) then yields%
\begin{equation}
\begin{array}{cl}
\mathbb{P}(a<\theta <b|\mathbf{y}) & \approx \exp \left\{ nf(\widehat{\theta 
})\right\} \int\limits_{a}^{b}\exp \left\{ -\frac{1}{2\sigma ^{2}}(\theta -%
\widehat{\theta })^{2}\right\} d\theta \\ 
& =\exp \left\{ nf(\widehat{\theta })\right\} \sqrt{2\pi \sigma ^{2}}\times
\{\Phi \lbrack \frac{b-\widehat{\theta }}{\sigma }]-\Phi \lbrack \frac{a-%
\widehat{\theta }}{\sigma }]\},%
\end{array}
\label{la}
\end{equation}%
where $\Phi (.)$ denotes the standard Normal cumulative distribution
function (cdf); and where, buried within the symbol `$\approx $' in (\ref{la}%
), is a rate of convergence that is a particular order of $n$, and is
probabilistic if randomness in $\mathbf{y}$ is acknowledged.

Not only did the result in (\ref{la}) represent the first step in the
development of Bayesian asymptotic \textit{theory}, it also provided a
simple practical solution to the \textit{computation }of general posterior
expectations like that in (\ref{gen_expect}). Two centuries later, \cite%
{tierney:kadane:1986} and \cite{tierney:kass:kadane:1989} revived and
formalized the Laplace approximation: using it to yield an asymptotic
approximation (of a given order) of any posterior expectation of the form of
(\ref{gen_expect}), including (in the multiple parameter case) marginal
posterior densities.

Two decades later, \cite{rue:martino:chopin:2009} then took the method
further: adapting it to approximate marginal posteriors (and {general
expectations like those in (\ref{gen_expect})}) in latent Gaussian models.
With the authors using a series of \textit{nested }Laplace approximations,
allied with low-dimensional numerical \textit{integration}, they termed
their method \textit{integrated nested Laplace approximation, }or INLA for
short. Since the latent Gaussian model class encompasses a large range of
empirically relevant models -- including, generalized linear models,
non-Gaussian state space (or hidden Markov) models, and spatial, or
spatio-temporal models -- a computational method tailored-made for such a
setting is sufficiently broad in its applicability to warrant detailed
consideration herein. In common with VB, and as follows from the use of
Laplace approximations evaluated at modal values, INLA eschews simulation
for optimization (in addition to using low-dimensional deterministic
integration methods).

Deferring to \cite{rue:martino:chopin:2009}, \cite{Rue2017}, \cite%
{martino2019integrated}, \cite{vanniekerk2019new} and \cite%
{wood2019simplified} for specific implementation details (including of the
latent Gaussian model structure), we provide here the \textit{key} steps of
INLA. To enhance the reader's understanding, we avoid the use of a summary
algorithmic presentation of the method. Consistent with our previous
notational convention, {we decompose }the full set of unknowns,\textbf{\ }$%
\boldsymbol{\theta }$, into an\textbf{\ }$m$-dimensional vector of
`hyperparameters' (in the language of INLA) that characterize the latent
Gaussian model,\textbf{\ }$\boldsymbol{\phi }$, and the full set of\textbf{\ 
}$K$ unknowns in the latent Gaussian field, denoted by\textbf{\ }$%
\boldsymbol{x}$. Each observation,{\ }$y_{i}$, $i=1,2,,,,n$, is assumed to
be independent, conditional on a linear predictor,{\ }$\eta _{i}$, {which is
modelled as a random function of }$\boldsymbol{x}$. {For computational
convenience, the vector} $\boldsymbol{\eta }=(\eta _{1},\eta _{2},...,\eta
_{n})^{\prime }$ {is} {also included as an element of }$\boldsymbol{x}${\ }({%
see \citealp{martino2019integrated}, for details).} The dimension, $K$, of $%
\boldsymbol{x}$ -- which contains observation-specific, plus common,
elements -- is larger, and potentially much larger, than the dimension of $%
\mathbf{y}$ itself. The model is then expressed as: 
\begin{equation}
\mathbf{y|x,}\boldsymbol{\phi }\sim \prod\nolimits_{i=1}^{n}p(y_{i}|\eta
_{i}(\mathbf{x}),\boldsymbol{\phi })\qquad {\ }\mathbf{x|}\boldsymbol{\phi }%
\sim \mathcal{N}(\boldsymbol{0},Q^{-1}(\boldsymbol{\phi }))\qquad {\ }%
\boldsymbol{\phi }\sim p(\boldsymbol{\phi }),  \label{GP_model}
\end{equation}%
where $Q(\boldsymbol{\phi })$ is the precision matrix of the latent Gaussian
field, assumed -- for computational feasibility -- to be sparse. The goal of
the authors is to approximate the marginal posteriors; $p(\phi _{j}|\mathbf{y%
})$, $j=1,2,..,m$, and $p(x_{k}|\mathbf{y})$, $k=1,2,..,K.$ The problems
envisaged are those in which $m$, the dimension of the hyperparameters $%
\boldsymbol{\phi }$, is small and $K$ is large (potentially in the order of
hundreds of thousands), with MCMC algorithms deemed to be {computationally
burdensome} as a consequence, due to the scale of the unknowns (and
potentially $\mathbf{y}$ also), and the challenging geometry of the
posterior. {We refer the reader to the references cited above for the wide
range of problems of this type to which INLA has been applied.}

Beginning with the expression of $p(\boldsymbol{\phi }|\mathbf{y})$ as%
\begin{equation}
p(\boldsymbol{\phi }|\mathbf{y})=\frac{p(\mathbf{x},\boldsymbol{\phi }|%
\mathbf{y})}{p(\mathbf{x}|\boldsymbol{\phi },\mathbf{y})}\propto \frac{p(%
\mathbf{x},\boldsymbol{\phi },\mathbf{y})}{p(\mathbf{x}|\boldsymbol{\phi },%
\mathbf{y})}=\frac{p(\mathbf{y|x,}\boldsymbol{\phi })p(\mathbf{x}|%
\boldsymbol{\phi })p(\boldsymbol{\phi })}{p(\mathbf{x}|\boldsymbol{\phi },%
\mathbf{y)}},  \label{theta}
\end{equation}%
and recognizing that the proportionality sign arises due to the usual lack
of integrating constant (over $\mathbf{x}$ and $\boldsymbol{\phi }$), the
steps of the method (in its simplest form) are as follows. First, on the
assumption that all components of the model can be evaluated and, hence,
that the numerator is available, $p(\boldsymbol{\phi }|\mathbf{y})$ in (\ref%
{theta}) is approximated as%
\begin{equation}
\widetilde{p}(\boldsymbol{\phi }|\mathbf{y})\propto \frac{p(\mathbf{y|%
\widehat{\mathbf{x}}}(\boldsymbol{\phi })\mathbf{,}\boldsymbol{\phi })p(%
\mathbf{\widehat{\mathbf{x}}}(\boldsymbol{\phi })|\boldsymbol{\phi })p(%
\boldsymbol{\phi })}{p_{G}(\mathbf{\widehat{\mathbf{x}}}(\boldsymbol{\phi })|%
\boldsymbol{\phi },\mathbf{y})}.  \label{approx_1}
\end{equation}%
The denominator in (\ref{approx_1}) represents a Gaussian approximation of $%
p(\mathbf{x}|\boldsymbol{\phi },\mathbf{y})$, $p_{G}(\mathbf{x}|\boldsymbol{%
\phi },\mathbf{y})=\mathcal{N}(\widehat{\mathbf{x}}(\boldsymbol{\phi }),%
\widehat{\Sigma }(\boldsymbol{\phi }))$, evaluated at the mode, $\widehat{%
\mathbf{x}}(\boldsymbol{\phi })$, of $p(\mathbf{x},\boldsymbol{\phi },%
\mathbf{y})$ (at a given value of $\boldsymbol{\phi }$), where $\widehat{%
\Sigma }(\boldsymbol{\phi })$ is the inverse of the Hessian of $-\log p(%
\mathbf{x},\boldsymbol{\phi },\mathbf{y)}$ with respect to $\mathbf{x}$,
also evaluated at $\widehat{\mathbf{x}}(\boldsymbol{\phi }).$ The expression
in (\ref{approx_1}) can obviously be further simplified to%
\begin{equation}
\widetilde{p}(\boldsymbol{\phi }|\mathbf{y})\propto p(\mathbf{y|\mathbf{%
\widehat{\mathbf{x}}(\boldsymbol{\phi }}),}\boldsymbol{\phi })p(\mathbf{%
\widehat{\mathbf{x}}(\boldsymbol{\phi }})|\boldsymbol{\phi }\mathbf{)}p(%
\boldsymbol{\phi })\left\vert \widehat{\Sigma }(\boldsymbol{\phi }%
)\right\vert ^{1/2},  \label{approx_2}
\end{equation}%
which, up to the integrating constant, is identical to the Laplace
approximation of a marginal density in Tierney and Kadane (1986, equation
(4.1)). \cite{rue:martino:chopin:2009} discuss the circumstances in which
the order of approximation proven in \cite{tierney:kadane:1986} applies to
the latent Gaussian model setting; whilst \cite{tang2021} provide further
approximation results pertaining to high-dimensional models.

With the marginal posterior for the $kth$ element of $\mathbf{x}$ defined as%
\begin{equation}
\widetilde{p}(x_{k}|\mathbf{y})=\int_{{\Theta }}\widetilde{p}(x_{k}|%
\boldsymbol{\phi },\mathbf{y)}\widetilde{p}(\boldsymbol{\phi }|\mathbf{y})d%
\boldsymbol{\phi },  \label{marg_zi}
\end{equation}%
a second application of a Laplace approximation would yield%
\begin{equation}
\widetilde{p}(x_{k}|\boldsymbol{\phi },\mathbf{y)}\propto p(\mathbf{y|%
\widehat{\mathbf{x}}}_{-k}\mathbf{(}\boldsymbol{\phi },x_{k})\mathbf{,}%
\boldsymbol{\phi })p(\mathbf{\widehat{\mathbf{x}}}_{-k}\mathbf{(}\boldsymbol{%
\phi },x_{k})|\boldsymbol{\phi }\mathbf{)}p(\boldsymbol{\phi })\left\vert 
\widehat{\Sigma }_{-k}(\boldsymbol{\phi },x_{k})\right\vert ^{1/2},
\label{approx_3}
\end{equation}%
where $\widehat{\mathbf{x}}_{-k}(\boldsymbol{\phi },x_{k})$ is the mode of $%
p(\mathbf{x}_{-k},x_{k},\boldsymbol{\phi },\mathbf{y})$ (at given values of $%
\boldsymbol{\phi }$ and $x_{k}$, with $\mathbf{x}_{-k}$ denoting all
elements of $\mathbf{x}$ other than the $kth$); and where $\widehat{\Sigma }%
_{-k}(\boldsymbol{\phi },x_{k})$ is the inverse of the Hessian of $-\log p(%
\mathbf{x}_{-k},x_{k},\boldsymbol{\phi },\mathbf{y)}$ with respect to $%
\mathbf{x}_{-k}$, also evaluated at $\widehat{\mathbf{x}}_{-k}(\boldsymbol{%
\phi },x_{k}).$ Computation of (\ref{approx_3}) for each $x_{k}$ would,
however, involve $K$ optimizations (over $\mathbf{x}_{-k}$) plus $K$
specifications of the high-dimensional matrix $\widehat{\Sigma }_{-k}(%
\boldsymbol{\phi },x_{k}).$ \cite{rue:martino:chopin:2009} avoid this
computational burden by modifying the approximation in (\ref{approx_3}) in a
number of alternative ways, all details of which are provided in the
references cited above. Once a representation of $\widetilde{p}(x_{k}|%
\boldsymbol{\phi },\mathbf{y)}$ is produced, (\ref{marg_zi}) is computed
using a deterministic numerical integration scheme defined over a grid of
values for the low-dimensional $\boldsymbol{\phi }.$

Defining the marginal posterior for the $jth$ element of $\boldsymbol{\phi }$
as $\widetilde{p}(\phi _{j}|\mathbf{y})=\int_{{\Theta }_{-j}}\widetilde{p}(%
\boldsymbol{\phi }|\mathbf{y})d\boldsymbol{\phi }_{-j},$ where $\boldsymbol{%
\phi }_{-j}$ denotes all elements of $\boldsymbol{\phi }$ excluding $\phi
_{j}$, this integral is computed using $m-$dimensional deterministic
integration over $\boldsymbol{\phi }_{-j}$, once again on the maintained
assumption that $m$ is small. Finally, if required, {the marginal likelihood}%
, $p(\mathbf{y})$ can be approximated by computing the normalizing constant
in (\ref{approx_2}), $\int_{{\Theta }}p(\mathbf{y|\mathbf{\widehat{\mathbf{x}%
}}}($\textbf{\textbf{$\boldsymbol{\phi }$}}$)\mathbf{,}\boldsymbol{\phi })p(%
\mathbf{\widehat{\mathbf{x}}}($\textbf{$\boldsymbol{\phi }$}$)|\boldsymbol{%
\phi }\mathbf{)}p(\boldsymbol{\phi })\left\vert \widehat{\Sigma }(%
\boldsymbol{\phi })\right\vert ^{1/2}d\boldsymbol{\phi },$ using
deterministic integration over $\boldsymbol{\phi }$.

All steps of the INLA algorithm can be implemented using the dedicated
package, R-INLA (available at www.r-inla.org), for the general LGM
framework, with particular packages also available for implementing INLA in
more specific models nested within the LGM class; see \cite%
{martino2019integrated} for a listing of all such packages.\ \cite{Gomez2018}
and \cite{berild2021importance} demonstrate how the INLA approach (and the
R-INLA software) can also be applied to models beyond the LGM class by means
of additional MCMC or IS\ sampling steps applied to models that are LGMs 
\textit{conditional} on certain fixed parameters. \cite{margossian2020} {%
extend} {INLA principles to the case in which }$m$ is too large{\ for
treatment by deterministic integration, by `embedding' INLA {within an HMC}
sampling scheme. In this case }$\widetilde{p}(\boldsymbol{\phi }|\mathbf{y})$
{- produced as in (\ref{approx_2}) - serves as the target density for the
HMC sampler, and each }$\widetilde{p}(\phi _{j}|\mathbf{y})$ {is estimated
via the HMC draws. Finally, }\cite{stringer2021} have adapted the standard
INLA methodology both to cater for an \textit{extended }class of LGMs, in
which the conditional independence assumption for $y_{i}$ is eschewed, and
to scale better to large data sets.

\subsubsection{Illustrative example: VB and INLA\label{ill_vb}}

We complete this section on approximate Bayesian inference via optimization
by displaying and discussing graphical output {from \cite%
{braun2010variational} and \cite{margossian2020}, in which, respectively, VB
and INLA are used to conduct inference. The selected illustration from %
\citeauthor{braun2010variational} highlights the feasibility, speed and
(comparable) predictive accuracy of VB, versus an MCMC comparator. The
illustration extracted from \citeauthor{margossian2020} compares the
accuracy and speed of the `embedded' HMC method with a `full' HMC algorithm,
in which \textit{both} the latent Gaussian field and the hyperparameters are
inferred via simulation.}

\paragraph{VB illustration\protect\bigskip \newline
}

{We record here certain output from }a particular simulation exercise in 
\cite{braun2010variational}, in which VB is used to perform inference on a
large-scale hierarchical model for consumer choice. This illustration shares
characteristics common to many applications of this approximate method (and,
indeed, of INLA too): \textit{1) }Very high-dimensional $\boldsymbol{\theta }%
\ $and $\mathbf{y}$; but, at the same time \textit{2) }An analytical
expression for the model,\textit{\ }$p(\mathbf{y}|\boldsymbol{\theta })$.

The model in question is a `random utility model' specified for $H$
customers, each with heterogeneous preferences or `tastes', and each having
to select from $J$ {items (or choices)}, each with $K$ {choice-specific}
attributes. The total number of unknowns comprise the $K$-dimensional
vectors of customer-specific preferences over the attributes, $\boldsymbol{%
\beta }_{h}$, which may be specific to each of the $h=1,2,...,H$ customers,
plus the mean vector ($\boldsymbol{\zeta }$) and variance-covariance matrix (%
$\boldsymbol{\Omega }$) of the $K$-dimensional Gaussian distribution that
models the distribution of preferences across the population. Hence, in
terms of our notation, the dimension of $\boldsymbol{\theta }$ is the
combined dimensions of $\boldsymbol{\zeta }$, $\boldsymbol{\Omega }$ and $%
\boldsymbol{\beta }_{1},...,\boldsymbol{\beta }_{H}.$ The vector of observed
data $\mathbf{y}$ comprises $T_{h}$ choice events across $H$ customers and
is thus of total length $H\ast T_{h}.$ A $(J\times K)$ matrix of observed
attributes encountered by {customer} $h=1,2,...,H$, at choice event $%
t=1,2,..,T_{h}$ completes the observed data, where we denote the full
(concatenated) matrix of observed attributes over agents and events simply
by $\mathbf{X}$. For the design scenario with the largest specifications, $%
H=25,000$, $T_{h}=25$, $J=12$ and $K=10.$

\begin{figure}[h]
\centering\includegraphics[height=12cm,width=18cm]{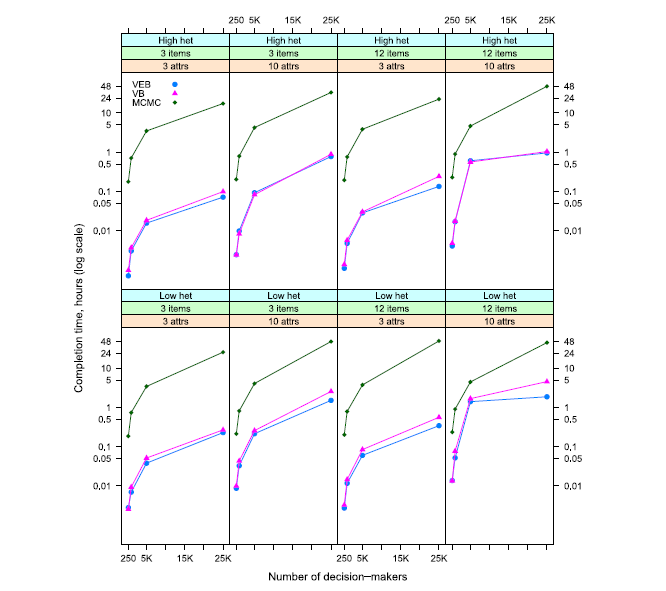}
\caption{Figure 2 from \protect\cite{braun2010variational} with caption:
``Timing results for Variational Empirical Bayes (VEB), Variational
Hierarchical Bayes (VB), and MCMC. Within each panel, completion time is
plotted on the log scale as a function of the number of agents, for fixed
values of the other simulation parameters (shown at the top of each panel).
In all simulated scenarios, variational methods complete more quickly than
MCMC. With 25,000 decision-makers, the variational algorithms complete in
five minutes to six hours, versus MCMC completion times of one to two days.
In the 25,000 agent case, the figure shows the time to generate 6000 MCMC
draws, based on a corresponding 1000-draw run (at which point the sampler
exhausted memory resources)."}
\label{fig2}
\end{figure}

A mean-field variational family $\mathcal{Q}$ is adopted, with a variational
approximation to $p(\boldsymbol{\beta }_{1},...,\boldsymbol{\beta }_{H},%
\boldsymbol{\zeta },\boldsymbol{\Omega }|\newline
\mathbf{X},\mathbf{y})$ chosen from $\mathcal{Q}$ to maximize the ELBO, via
a block coordinate ascent algorithm implemented with analytical expressions
for the gradient and the Hessian of the criterion function (%
\citealp{braun2010variational}, Appendix A). Whilst MCMC is obviously
challenging for this particular model, due to the scale of both $\boldsymbol{%
\theta }$ and $\mathbf{y}$, and, indeed, exhausts machine memory at a very
small number of iterations ($1000$), it \textit{is }feasible; hence, one aim
of this simulation exercise is to illustrate the \textit{relative speed }of
VB versus MCMC, where the MCMC algorithm is that of \cite{rossi2003}. We
display (as our own Figure \ref{fig2}) Figure 2 from \cite%
{braun2010variational}, retaining the original caption as, in tandem with
the explanatory material above, it is sufficiently informative to allow the
results to be interpreted without access to the paper. We note that in the
body of the figure: `items' refers to $J;$ `attrs' refers to $K;$ and
`Low/High het' refers to magnitude of the diagonal elements of $\boldsymbol{%
\Omega }$ (i.e. the degree of heterogeneity in the preferences of the
customer population). In the key, `VB' refers to the method summarized
herein, and `VEB' to the use of VB to implement empirical Bayes (which we do
not discuss here, for reasons of space)

\medskip As in Section \ref{ill_opt}, we highlight the key messages to be
gleaned from\textbf{\ }Figure \ref{fig2}:\medskip

\begin{enumerate}
\item For the scenario with $H=25,000$, $T_{h}=25$, $J=12$ and $K=10$ and
high heterogeneity (top right-hand panel), MCMC uses two days of computation
time to produce $6,000$ iterations, versus one hour for VB. In the same
setting, but with low heterogeneity (bottom right-hand panel), the
comparison is two days versus 6 hours. That is, VB is between 8 and 48 times
faster than MCMC.

\item For these two large-scale scenarios for $J$ and $K$, as $H$ (plotted
on the horizontal axis) increases, VB also scales noticeably better to the
consequent increase in $\mathbf{y}$ than does MCMC (i.e. the VB plots
flatten more than do the MCMC plots).

\item Similar comparable relativities between the MCMC and VB computational
burdens obtain for all other scenarios, although the superior scaling
performance of VB is less noticeable.

\item One would expect the use of an SVI algorithm to greatly reduce the
time taken to tackle the largest versions of the problem, and hence render
the performance gains of VB over MCMC even more marked.

\item In addition to the speed comparison documented in Figure \ref{fig2},
the authors report (\citealp{braun2010variational}, Appendix A, Tables 1 and
2) {that the accuracy with which the VB- and MCMC-based predictives match
the true predictive choice distribution (known in this artificial data
setting, and defined for an `average agent' and a `typical' item attribute)
is almost identical. }This result tallies with subsequent results in the VB
literature (see, e.g. \citealp{quiroz2018gaussian},\ and %
\citealp{frazierloss}), which demonstrate that predictive results obtained
using VB are largely unaffected by the inferential inaccuracy of the VB
posterior approximation.
\end{enumerate}

\paragraph{INLA illustration\protect\bigskip \newline
}

As a final illustration we report selected results from \cite{margossian2020}%
, in which a combination of INLA and HMC (referred to by the authors as the
`embedded' Laplace approximation) is applied to a spatial model for
mortality counts in Finland. In brief, conditionally Poisson mortality
counts ($y_{i}$), aggregated over 100 geographical regions ($i=1,2,...,n=100$%
), are modelled using a latent Gaussian process. Whilst the overarching aim
of {\citeauthor{margossian2020}} is to adapt INLA to cater for a very
high-dimensional hyperparameter vector ($\boldsymbol{\phi }$), and whilst
INLA itself was developed for the case of a high-dimensional latent Gaussian
field ($\mathbf{x}$), this illustrative example aims to compare the speed
and accuracy of the embedded method with that of a full HMC algorithm;
hence, both\textbf{\ }$\boldsymbol{\phi }$ and $\mathbf{x}$ are very
low-dimensional. Specifically, for each region $i$, $y_{i}|\eta _{i}\sim
Poisson(y_{e}^{i}\exp (\eta _{i}))$, where $y_{e}^{i}$ is the standardized
expected number of deaths, and $\eta _{i}$ is a linear function of a
two-dimensional vector of regional characteristics, $\mathbf{x}_{i}$. An
exponentiated quadratic kernel defines the elements of $Q^{-1}(\boldsymbol{%
\phi })$ in (\ref{GP_model}), and the two-dimensional vector $\boldsymbol{%
\phi }$ comprises the standard deviation ($\alpha $) and length scale ($l$)
in the kernel function. (See \citealp{Vanhatalo2010}, for all details of the
general model structure in which the specification used by {%
\citeauthor{margossian2020}} is nested.)

\begin{figure}[h]
\centering\includegraphics[height=8cm,width=14cm]{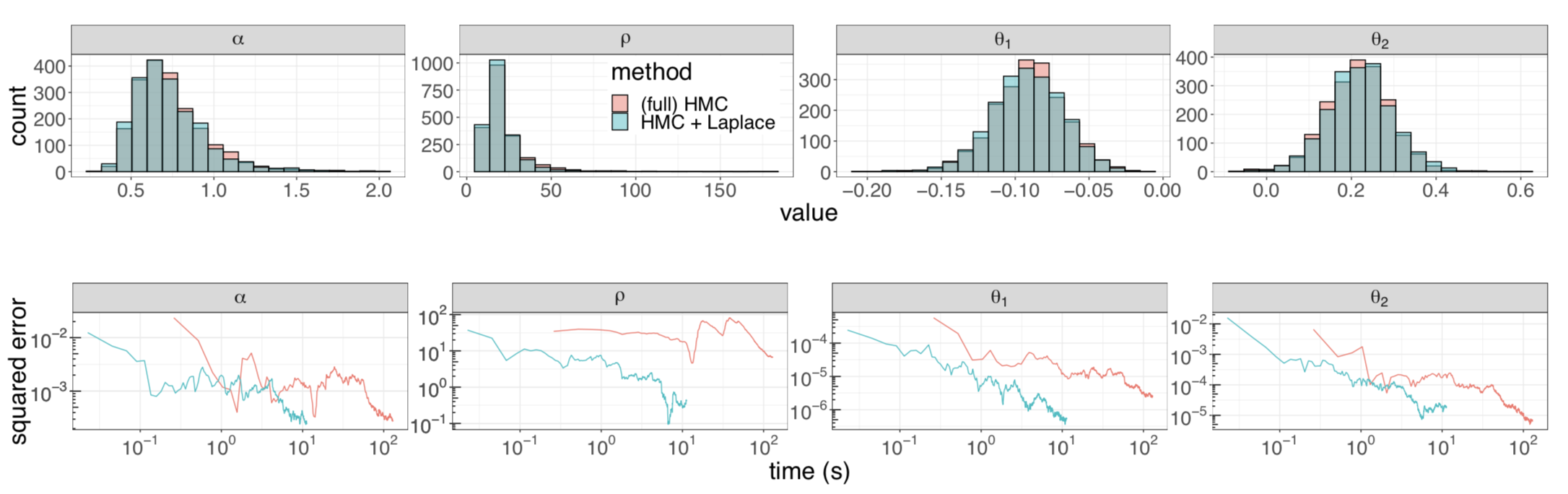}
\caption{Figure 2 from \protect\cite{margossian2020} with caption:
\textquotedblleft (Up) Posterior samples obtained with full HMC and the
embedded Laplace approximation when fitting the disease map. (Down) Error
when estimating the expectation value against wall time. Unreported in the
figure is that we had to fit full HMC twice before obtaining good tuning
parameters."}
\label{fig3}
\end{figure}

We display (as our own Figure \ref{fig3}) Figure 2 from \cite{margossian2020}%
, including the original caption, which is sufficiently informative. We do
note, however, that the authors use the notation $\theta _{i}$ to denote $%
\eta _{i}$, and they record -- in addition to results for $\alpha $ and $l$
-- results for the first two elements, $\theta _{1}$ ($\equiv \eta _{1}$)
and $\theta _{2}$ ($\equiv \eta _{2}$). 

The key highlights of\textbf{\ }Figure \ref{fig3} are as follows:

\begin{enumerate}
\item For this example the marginal posteriors for the four unknowns
(plotted in the top panel) produced by both the embedded Laplace
approximation and the full HMC algorithm are very similar; with both
algorithms based on 500 burn-in iterates and 500 subsequent draws.

\item At the same time, as documented in the bottom panel, the speed with
which the embedded approach estimates the relevant posterior expectations
produced from 18,000 HMC draws, to a certain level of precision, is an order
of magnitude greater than the full HMC method.

\item Finally, on the matter of speed, the authors illustrate that the speed
gains of the embedded method can be case-specific, depending, in part, on
the relative dimensions of $\boldsymbol{\phi }$\ and $\mathbf{x}$; and with
particular gains to be had when the dimension of $\mathbf{x}$ is much larger
than that of $\boldsymbol{\phi }$. Nevertheless, the authors do highlight
that, even without dramatic speed gains, the use of INLA to integrate out $%
\mathbf{x}$ does avoid the delicate tuning required to implement HMC
successfully in such a high-dimensional space.
\end{enumerate}

\subsection{Hybrid approximate methods\label{hybrid}}

\subsubsection{Overview}

We remind the reader at this point of the following:\textit{\ i)} whilst ABC
and BSL are advantageous when $p(\mathbf{y}|\boldsymbol{\theta })$ cannot be
evaluated, a large dimension for $\boldsymbol{\theta }$ (and, hence, for $%
\eta (\mathbf{y})$) causes challenges (albeit to differing degrees) for
both;\ \textit{ii)} VB and INLA are much better equipped to deal with
high-dimensional $\boldsymbol{\theta }$ (and/or $\mathbf{y}$)$,$ but require
the evaluation of $p(\boldsymbol{\theta },\mathbf{y})$ and, thus, $p(\mathbf{%
y}|\boldsymbol{\theta })$. Recently, hybrid algorithms that meld aspects of
ABC, BSL and VB, along with so-called \textit{pseudo-marginal} principles,
have been used to deal with settings in which the likelihood is intractable 
\textit{and} either $\boldsymbol{\theta }$ or $\mathbf{y}$, or both, are
high-dimensional. A hybrid method that reduces the impact of dimensionality
on ABC by introducing Gibbs steps has also been proposed. All such `mixed'
techniques are outlined below, after a very brief outline of pseudo-marginal
MCMC.

\subsubsection{A brief introduction to pseudo-marginal MCMC}

\label{sec:psudo}

The (combined) insight of \cite{beaumont:2003} and \cite%
{andrieu:roberts:2009} began with the following observation. Use $\mathbf{%
u\in }$ $\mathcal{U}$ to denote all of the canonical (problem-specific)
random variables that may be used to produce an \textit{unbiased }estimate
of the likelihood function, $p(\mathbf{y}|\boldsymbol{\theta })$. Using the
now standard concept of `data augmentation' (\citealp{tanner87}), an MCMC
scheme can then be applied to the joint space $(\boldsymbol{\theta }\mathbf{%
,u})$, in order to target the required invariant distribution, $p(%
\boldsymbol{\theta }\mathbf{|y}).$ An informal demonstration of this result
is straightforward. Define $h(\mathbf{u})$ as the distribution of $\mathbf{u}
$ (independently of the prior $p(\boldsymbol{\theta })$), and let $h(\mathbf{%
y}|\boldsymbol{\theta }\mathbf{,u})$ denote an estimate of the likelihood $p(%
\mathbf{y}|\boldsymbol{\theta })$, that is unbiased in the sense that $E_{%
\mathbf{u}}[h(\mathbf{y}|\boldsymbol{\theta },\mathbf{u})]=p(\mathbf{y}|%
\boldsymbol{\theta })$. Then we have that $h(\boldsymbol{\theta }\mathbf{|y}%
)\propto \int\nolimits_{\mathcal{U}}h(\mathbf{y}|\boldsymbol{\theta }\mathbf{%
,u})p(\boldsymbol{\theta })h(\mathbf{u})d\mathbf{u}=p(\boldsymbol{\theta }%
)E_{\mathbf{u}}[h(\mathbf{y}|\boldsymbol{\theta },\mathbf{u})]=p(\boldsymbol{%
\theta })p(\mathbf{y}|\boldsymbol{\theta })\propto p(\boldsymbol{\theta }%
\mathbf{|y})$, as desired. That is, in yielding a chain with an invariant
distribution equal to the correct marginal, $p(\boldsymbol{\theta }\mathbf{|y%
})$, a pseudo-marginal method produces an exact simulation-based estimate of
(\ref{gen_expect}).

Application of the pseudo-marginal principle to a Metropolis-Hastings MCMC
algorithm involves substituting $h(\mathbf{y}|\boldsymbol{\theta }\mathbf{,u}%
)$ for $p(\mathbf{y}|\boldsymbol{\theta })$ in the expression defining the
acceptance probability (\citealp{chib:greenberg:1995}), with the term
pseudo-marginal Metropolis-Hastings used in this case. When the unbiased
likelihood estimate is produced specifically via the use of \textit{particle
filtering} in a state space model, the term \textit{particle} MCMC (PMCMC)
has also been coined \citep{andrieu:doucet:holenstein:2010}.

Pseudo-marginal principles play a role in the hybrid methods in Sections \ref%
{hy1} and \ref{hy3} below.\footnote{%
We refer to \cite{doucet2015efficient}, \cite{deligiannidis2018correlated}, 
\cite{bardenet2017markov}%
, 
\cite{quiroz2018speeding}%
, 
\cite{quiroz2019speeding}
and 
\cite{moores2020bayesian}
for applications of pseudo-marginal MCMC methods (in their own right) to
intractable problems.}

\subsubsection{VB with intractable likelihoods\label{hy1}}

\cite{tran2017variational} devise a hybrid VB/pseudo-marginal method for use
when the likelihood function is intractable, coining the technique `VBIL'.
To appreciate the principles of the method, consider that the variational
approximation is indexed by a finite dimensional parameter $\boldsymbol{%
\lambda }$, so that $\mathcal{Q}:=\{\boldsymbol{\lambda }\in \Lambda :q_{%
\boldsymbol{\lambda }}\}$. The variational approximation is then obtained by
maximizing the ELBO, $\mathcal{L}(\boldsymbol{\lambda }):=\text{ELBO}[q_{%
\boldsymbol{\lambda }}(\boldsymbol{\theta })]$, over $\Lambda $. VBIL
replaces the intractable likelihood $p(\mathbf{y}|\boldsymbol{\theta })$
with an estimator $h(\mathbf{y}|\boldsymbol{\theta },\mathbf{u})$, such that 
$E_{\mathbf{u}}[h(\mathbf{y}|\boldsymbol{\theta },\mathbf{u})]=p(\mathbf{y}|%
\boldsymbol{\theta })$, and considers as target distribution the joint
posterior 
\begin{equation*}
h(\boldsymbol{\theta },z|\mathbf{y})\propto \pi (\boldsymbol{\theta })h(%
\mathbf{y}|\boldsymbol{\theta },\mathbf{u})\exp (z)g(z|\boldsymbol{\theta }),%
\text{ where }z:=\log h(\mathbf{y}|\boldsymbol{\theta },\mathbf{u})-\log p(%
\mathbf{y}|\boldsymbol{\theta }),
\end{equation*}%
and where $g(z|\boldsymbol{\theta })$ denotes the distribution of $z|%
\boldsymbol{\theta }$. Given that $h(\mathbf{y}|\boldsymbol{\theta },\mathbf{%
u})$ is, by construction, an unbiased estimator of $p(\mathbf{y}|\boldsymbol{%
\theta })$, it follows that marginalizing over $z$ in $h(\boldsymbol{\theta }%
,z|\mathbf{y})$, yields the posterior distribution of interest, namely $p(%
\boldsymbol{\theta }|\mathbf{y})$. \citeauthor{tran2017variational} then
minimize $\text{KL}[q_{\lambda }(\boldsymbol{\theta },z)|p(\boldsymbol{%
\theta },z|\mathbf{y})]$ over\textbf{\ }the augmented space of $(\boldsymbol{%
\theta },z)$, using as the variational family $\mathcal{Q}$ distributions of
the form $q_{\boldsymbol{\lambda }}(\boldsymbol{\theta },z)=q_{\boldsymbol{%
\lambda }}(\boldsymbol{\theta })g(z|\boldsymbol{\theta })$. Whilst, in
general, minimization of $\text{KL}[q_{\boldsymbol{\lambda }}(\boldsymbol{%
\theta },z)|p(\boldsymbol{\theta },z|\mathbf{y})]$ is not the same as
minimization of $\mathcal{L}(\boldsymbol{\lambda })$, the authors
demonstrate the two solutions \textit{do} correspond under particular tuning
regimes for $h(\mathbf{y}|\boldsymbol{\theta },\mathbf{u})$.

Following \cite{tran2017variational}, \cite{ong2018variational} propose an
alternative VB-based method for intractable likelihood problems. The authors
begin with the recognition that establishing the conditions under which the
minimizers of $\text{KL}[q_{\boldsymbol{\lambda }}(\boldsymbol{\theta },z)|p(%
\boldsymbol{\theta },z|\mathbf{y})]$ and $\mathcal{L}(\boldsymbol{\lambda })$
coincide is non-trivial, and that in certain types of problems it may be
difficult to appropriately tune $h(\mathbf{y}|\boldsymbol{\theta },\mathbf{u}%
)$ so that they coincide. This acknowledgement then prompts them to
construct a variational approximation of a simpler target, namely the BSL
posterior in \eqref{bsl_1}. By focusing on the (simpler) approximate
posterior, rather than the exact posterior $p(\boldsymbol{\theta }|\mathbf{y}%
)$, the approach of \citeauthor{tran2017variational} can be recycled using
any unbiased estimator of the synthetic likelihood, $p_{a}(\eta (\mathbf{y})|%
\boldsymbol{\theta })$ -- which we recall is nothing but a Normal likelihood
with unknown mean and variance-covariance matrix -- of which several
closed-form examples exist. Moreover, since the approach of %
\citeauthor{ong2018variational} does not rely on the random variables $%
\mathbf{u}$ in order for its likelihood estimate to be unbiased, no tuning
of $h(\mathbf{y}|\boldsymbol{\theta },\mathbf{u})$ is required, and the
minimizers of $\text{KL}[q_{\boldsymbol{\lambda }}(\boldsymbol{\theta },z)|p(%
\boldsymbol{\theta },z|\mathbf{y})]$ and $\mathcal{L}(\boldsymbol{\lambda })$
will always coincide.

While useful, it must be remembered that the approach of \cite%
{ong2018variational} targets only the \textit{partial} posterior $p(%
\boldsymbol{\theta }|\eta (\mathbf{y}))$. Furthermore, given the discussion
in Section \ref{bsl}, the approach is likely to perform poorly when the
summaries used to construct the unbiased estimator of the synthetic
likelihood $p_{a}(\eta (\mathbf{y})|\boldsymbol{\theta })$ are non-Gaussian.
Given that, by definition, the problem is a high-dimensional one, thereby
requiring a large collection of summaries, the Gaussian approximation for $%
\eta (\mathbf{y})$ may not be accurate.

\subsubsection{VB and ABC\label{hy2}}

Similar to the above, \cite{barthelme:chopin:2014} and \cite%
{barthelme2018divide} propose the use of variational methods to approximate
the ABC posterior. The approach of \citeauthor{barthelme:chopin:2014} is
based on `local' collections of summary statistics that are computed by
first partitioning the data into $b\leq n$ distinct `chunks', $\mathbf{y}%
_{1},\dots ,\mathbf{y}_{b}$, with possibly differing lengths and support,
and then computing the summaries $\eta (\mathbf{y}_{i})$ for each of the $b$
chunks. Using this collection of local summaries, the authors then seek to
compute an approximation to the following ABC posterior: 
\begin{equation}
p_{\varepsilon }(\boldsymbol{\theta }|\eta (\mathbf{y}))\propto p(%
\boldsymbol{\boldsymbol{\theta }})\prod_{i=1}^{b}\left\{ \int p(\mathbf{z}%
_{i}|\mathbf{y}_{1:i-1},\boldsymbol{\boldsymbol{\theta }})\mathbb{I}\left\{
\left\Vert \eta \left( \mathbf{z}_{i}\right) -\eta \left( \mathbf{y}%
_{i}\right) \right\Vert \leq \varepsilon \right\} d\mathbf{z}_{i}\right\} =p(%
\boldsymbol{\boldsymbol{\theta }})\prod_{i=1}^{b}\ell _{i}(\boldsymbol{%
\boldsymbol{\theta }}),  \label{abc_st}
\end{equation}%
which implicitly maintains that the `likelihood chunks', $\ell _{i}(%
\boldsymbol{\boldsymbol{\theta }})$, $i=1,2,...,b,$ are conditionally
independent.

The posterior in (\ref{abc_st}) is then approximated using expectation
propagation (EP) (see \citealp{bishop2006pattern}, Chapter 10, for details).
The EP approximation seeks to find a tractable density $q_{\boldsymbol{%
\lambda }}(\boldsymbol{\theta })\in \mathcal{Q}$ that is close to $%
p_{\varepsilon }(\boldsymbol{\theta }|\eta (\mathbf{y}))$ by minimizing $%
\text{KL}\left[ p_{\varepsilon }(\boldsymbol{\theta }|\eta (\mathbf{y}))|q_{%
\boldsymbol{\lambda }}(\boldsymbol{\theta })\right] $. The reader will note
that this minimization problem is actually the reverse of the standard
variational problem in \eqref{vb_1}, and is a feasible variational problem
because $p_{\varepsilon }(\boldsymbol{\theta }|\eta (\mathbf{y}))$ is
accessible. Using a factorizable Gaussian variational family with
chunk-specific {mean} {vector} and {covariance matrix}, $\boldsymbol{\mu }%
_{i}$ and $\boldsymbol{\Sigma }_{i}$, respectively, $i=1,\dots ,b$, i.e., $%
\mathcal{Q}:=\{\boldsymbol{\lambda }=(\boldsymbol{\lambda }_{1},\dots ,%
\boldsymbol{\lambda }_{b})\in \Lambda :q_{\boldsymbol{\lambda }}(\boldsymbol{%
\theta }):=\prod_{i=1}^{b}q_{i,\boldsymbol{\lambda }_{i}}(\boldsymbol{\theta 
})\}$, this minimization problem is solved iteratively by minimizing the KL
divergence between $\ell _{i}(\boldsymbol{\theta })$ and $q_{i,\boldsymbol{%
\lambda }_{i}}(\boldsymbol{\theta })$ for $i=1,2,...,b$. A coordinate ascent
optimization approach allows the $i$-th variational component to be updated
by calculating (using Monte Carlo integration) the {mean vector and
covariance matrix} of $q_{i,\boldsymbol{\lambda }_{i}}(\boldsymbol{\theta })$%
, based on data simulated from $\ell _{i}(\boldsymbol{\theta })$,
conditional on $\boldsymbol{\theta }$ drawn from the variational
approximation based on the remaining $j\neq i$ chunks.

By chunking data to create conditionally independent likelihood increments,
and by employing (conditionally independent) Gaussian approximations over
these chunks, EP-ABC creates a (sequentially updated) Gaussian
pseudo-posterior that serves as an approximation to the original ABC
posterior. Given that EP-ABC requires the posterior approximation to be
Gaussian (or more generally within the linear exponential family), the
resulting EP-ABC posterior may not be a reliable approximation to the ABC
posterior if the data has strong, or nonlinear, dependence, or (similar to
the problem identified for BSL) if (\ref{abc_st}) has non-Gaussian features,
such as thick tails, multimodality or boundary issues. Moreover, the need to
generate synthetic data sequentially according to different chunks\ of the
likelihood is unlikely to be feasible in models where there is strong or
even moderate serial dependence, and generation of new data requires
simulating the entire path history up to that point.

\subsubsection{ABC and Gibbs sampling\label{Gibbs}}

As described in Section \ref{summ}, ABC methods suffer from a curse of
dimensionality. Whilst this is typically expressed in terms of the
dimensionality of the summary statistics, there is obviously an intrinsic
link between the dimension of the summaries and that of $\boldsymbol{\theta }
$ itself, with the dimension of $\boldsymbol{\theta }$ necessarily imposing
a lower bound on the dimension of\textbf{\ $\eta \left( \mathbf{y}\right) $}
that can be used to guarantee identification (see \citealp{FMRR2016}). \cite%
{Nott2014} and \cite{martin2019auxiliary} both provide ways of alleviating
this particular issue by advocating a separate selection process for
individual elements (or blocks) of $\boldsymbol{\theta }$, with
corresponding elements (or blocks) of $\eta \left( \mathbf{y}\right) $ used
in the matching process. Different criteria are used in each paper to define
what is meant by `corresponding'. This component-by-component approach is
shown to produce more effective algorithms, within the confines of the
particular examples explored, but no formal investigation or validation of
the principle is undertaken in either piece of work.

\cite{clarte2019component} have attempted to formalize the approach, by
combining the principles of ABC and Gibbs sampling, which they refer to as
ABC-Gibbs (see also \citealp{Kousathanas2016}, and %
\citealp{rodrigues2019likelihood}, for related work). In short, the vector $%
\boldsymbol{\theta }$ is blocked in a suitable way, and conditional
posteriors defined as in a standard Gibbs sampling algorithm. Simulation
from each posterior then occurs via an ABC step (along the lines of
Algorithm \ref{ABCalg2}, for example) but with a summary statistic chosen to
be informative about the component of $\boldsymbol{\theta }$ that is the
argument of the conditional posterior - a choice that is deemed to be easier
than choosing informative summaries about the full $\boldsymbol{\theta }.$
In the case where a conditional can be simulated from directly, the
approximation step is not required.

While questions remain regarding the theoretical behaviour of the hybrid
algorithm, the authors do establish some sufficient conditions for the
convergence of the algorithm, with convergence being to the limiting
distribution of reject/accept ABC in certain cases. They also demonstrate
notable improvement in the numerical efficiency of the algorithm, in
comparison with both reject/accept ABC and a particular ABC-SMC method.

\subsubsection{ABC and PMCMC\label{hy3}}

Thus far we have discussed the use of ABC to conduct direct inference on the
fixed, static or global parameters. In state space settings, in which both
global and local parameters (or latent states) feature, ABC principles have
also been used to the implement the particle filtering that is often
required as an intermediate step towards conducting inference on the global
unknowns. This has been particularly useful in cases where the measurement
density has no closed form and, hence, cannot be used to define the particle
weights in the usual way. In this case, the matching principle that
underpins ABC is applied at a single observation level, one time point at a
time, and without summarization if the data is one-dimensional. This process
of `ABC filtering' then provides a simulation-based estimate of the
likelihood function. This can, in turn, be used either as a basis for
producing frequentist point estimates of the parameters\ (%
\citealp{jasra:etal:2012}; \citealp{calvet2015accurate}) or -- in the spirit
of this section on Bayesian hybrids -- as an input into a PMCMC scheme (%
\citealp{dean2014parameter}; \citealp{jasra2015approximate}).

\section{Future Directions for Approximate Methods\label{future}}

We end our review of approximate Bayesian methods by documenting work that
addresses the following questions: \textit{1)} What are the implications for
approximate computation if an assumed parametric model is misspecified?; 
\textit{2)} What are the implications for approximate computation if the
conventional likelihood-based paradigm is eschewed altogether, and a \textit{%
generalized, robust }Bayesian, or \textit{moment-based} approach to
inference (e.g. \citealp{bissiri:etal:2016}; \citealp{chib2018bayesian}; %
\citealp{miller2019robust}; \citealp{loaiza2019focused}) is adopted?; and 
\textit{3) }What role can approximate computation play in Bayesian
prediction?

\begin{enumerate}
\item[\textit{1)}] Papers that address the first question (with reference to
ABC, BSL and VB respectively) are as follows. \textit{First}, \cite%
{frazier2020model} analyze the theoretical properties of ABC under model
misspecification; outlining when ABC concentrates posterior mass on an
appropriately defined pseudo-true value, and when it does not. The
nonstandard asymptotic behaviour of the ABC posterior, including its failure
to yield credible sets with valid frequentist coverage, is highlighted. The
authors also devise techniques for diagnosing model misspecification in the
context of ABC. \textit{Second, }similar to ABC, \cite{frazier2021synthetic}
demonstrate that BSL displays non-standard behavior under model
misspecification: depending on the nature and level of model
misspecification, the BSL posterior may be approximately Gaussian,
mixed-Gaussian, or concentrate onto the boundary of the parameter space. In
a similar vein, \cite{frazier2019robust} devise a version of BSL that is
robust to model misspecification, and demonstrate that this version can be
much more computationally efficient than standard BSL when the model is
misspecified. \textit{Third}, \cite{alquier2020concentration} and \cite%
{zhang2017convergence} investigate posterior concentration of VB methods
under model misspecification. Both pairs of authors demonstrate that the VB
posterior concentrates onto the value that minimizes the Kullback-Leibler
(KL) divergence from the true DGP.

\item[\textit{2) }] With reference to the second question, \cite%
{knoblauch2019generalized} propose what they term \textit{generalized
variational inference, }by extending the specification of the Bayesian
paradigm to accommodate general loss functions (thereby avoiding the
reliance on potentially misspecified likelihoods) and building an VB
computational tool within that setting. {In a somewhat similar spirit, \cite%
{schmon2021generalized}} extend ABC to accommodate general loss functions,
with \cite{pacchiardi2021generalized} applying a similar approach to deal
with intractable likelihoods in the context of scoring rules.\medskip 
\newline
\cite{frazierloss} also apply an approximate method in a setting in which a
general loss function is specified, {but }with predictive accuracy dictating
the form of the {loss. A VB approximation} of the resultant `Gibbs
posterior' (\citealp{Zhang2006a}; \citealp{Zhang2006b}; \citealp{jiang2008}) 
{is adopted} {due} the high dimensionality of the {problems tackled}. The
authors prove theoretically, and illustrate numerically, {that for a large
enough value of }$n$ there is no reduction in predictive accuracy as a
result of approximating the posterior via VB. \medskip \newline
Finally, \cite{tran2019variational} extend VB to manifolds, rather than
using VB to approximate the conventional likelihood-based posterior.

\item[\textit{3) }] {The work by \cite{frazierloss}} {cited above }continues
in the vein of other work in which approximate computation plays a role in
Bayesian prediction. \cite{frazier2019approximate}, for instance, produce an
approximation of $p(y_{n+1}^{\ast }|\mathbf{y})$ in (\ref{predict}) by using
an ABC-based posterior to replace $p(\boldsymbol{\theta }\mathbf{|y})$. The
approximate predictive is numerically indistinguishable from the exact
predictive (in the cases investigated), and yields equivalent out-of-sample
accuracy as a consequence. Further, under the regularity that ensures
Bayesian consistency for both the exact and ABC posteriors, the exact and
approximate predictives are shown to be asymptotically equivalent. (See also %
\citealp{canale2016}, and \citealp{konkamking2019}.) Other work produces an
approximate predictive by using a VB approximation of the (likelihood-based)
posterior (\citealp{tran2017variational}; \citealp{quiroz2018gaussian}; %
\citealp{koop2018variational}; \citealp{chan2020fast}; %
\citealp{loaiza2020fast}). The tenor of this work is somewhat similar to
that of \cite{frazier2019approximate} and \cite{frazierloss}; that is,
computing the posterior via an approximate method does not necessarily
reduce predictive accuracy. In contrast, \cite{frazier2021note} document an
important case where the approximation \textit{can} matter. In brief, the
use of a VB approximation to the posterior of the{\ local parameters} in a
state space model \textit{is }found to impinge on predictive accuracy in
some cases, due to the lack of Bayesian consistency of the posterior for the 
{global parameters} that can arise.
\end{enumerate}

In summary, approximate Bayesian methods are beginning to confront -- and
adapt to -- the reality of misspecified DGPs, and the generalizations beyond
the standard likelihood-based update that are {increasingly adopted}. Their
good performance in many predictive settings is also encouraging. Being able
to tackle intractable problems via an approximate method without
compromising predictive accuracy is an attractive prospect for
investigators, and suggests that approximate computation may play an
increasingly large role in complex predictive settings, over and above its
critical role in inference.

{\footnotesize \ \baselineskip8pt \setlength{\bibsep}{8pt} 
\bibliographystyle{apalikeit}
\bibliography{Bayes_comp}

\begin{thebibliography}{}

\bibitem[Alquier and Ridgway, 2020]{alquier2020concentration}
Alquier, P. and Ridgway, J. (2020).
\newblock Concentration of tempered posteriors and of their variational
  approximations.
\newblock {\em The Annals of Statistics}, 48(3):1475--1497.

\bibitem[An \emph{et~al.}, 2019]{an2019bsl}
An, Z., South, L.~F., and Drovandi, C. (2019).
\newblock {BSL}: An {R} package for efficient parameter estimation for
  simulation-based models via {B}ayesian synthetic likelihood.

\bibitem[Andrieu \emph{et~al.}, 2011]{andrieu:doucet:holenstein:2010}
Andrieu, C., Doucet, A., and Holenstein, R. (2011).
\newblock Particle {M}arkov chain {M}onte {C}arlo.
\newblock {\em J. Royal Statist. Society Series B}, 72(2):269--342.
\newblock With discussion.

\bibitem[Andrieu and Roberts, 2009]{andrieu:roberts:2009}
Andrieu, C. and Roberts, G. (2009).
\newblock The pseudo-marginal approach for efficient {M}onte {C}arlo
  computations.
\newblock {\em Ann. Statist.}, 37(2):697--725.

\bibitem[Ardia \emph{et~al.}, 2012]{ARDIA2012}
Ardia, D., Baştürk, N., Hoogerheide, L., and van Dijk, H.~K. (2012).
\newblock A comparative study of {M}onte {C}arlo methods for efficient
  evaluation of marginal likelihood.
\newblock {\em Computational Statistics and Data Analysis}, 56(11):3398--3414.

\bibitem[Bardenet \emph{et~al.}, 2017]{bardenet2017markov}
Bardenet, R., Doucet, A., and Holmes, C. (2017).
\newblock On {M}arkov chain {M}onte {C}arlo methods for tall data.
\newblock {\em J. Machine Learning Res.}, 18(1):1515--1557.

\bibitem[Barthelm{\'e} \emph{et~al.}, 2018]{barthelme2018divide}
Barthelm{\'e}, S., Chopin, N., and Cottet, V. (2018).
\newblock Divide and conquer in {ABC}: Expectation-propagation algorithms for
  likelihood-free inference.
\newblock {\em Handbook of Approximate {B}ayesian Computation}, pages 415--34.
\newblock Chapman \& Hall/CRC. Eds. Sisson, S., Fan, Y., Beaumont, M.

\bibitem[Barthelmé and Chopin, 2014]{barthelme:chopin:2014}
Barthelmé, S. and Chopin, N. (2014).
\newblock Expectation propagation for likelihood-free inference.
\newblock {\em J. American Statist. Assoc.}, 109(505):315--333.

\bibitem[Bauwens and Richard, 1985]{bauwens:1985}
Bauwens, L. and Richard, J. (1985).
\newblock A 1-1 {P}oly-$t$ random variable generator with application to
  {M}onte {C}arlo integration.
\newblock {\em J. Econometrics}, 29(1):19--46.

\bibitem[Beaumont, 2003]{beaumont:2003}
Beaumont, M. (2003).
\newblock Estimation of population growth or decline in genetically monitored
  populations.
\newblock {\em Genetics}, 164:1139--1160.

\bibitem[Beaumont, 2010]{beaumont:2010}
Beaumont, M. (2010).
\newblock Approximate {B}ayesian computation in evolution and ecology.
\newblock {\em Annual Review of Ecology, Evolution, and Systematics},
  41:379--406.

\bibitem[Beaumont \emph{et~al.}, 2009]{beaumont:cornuet:marin:robert:2009}
Beaumont, M., Cornuet, J.-M., Marin, J.-M., and Robert, C. (2009).
\newblock Adaptive approximate {B}ayesian computation.
\newblock {\em Biometrika}, 96(4):983--990.

\bibitem[Beaumont \emph{et~al.}, 2002]{be02}
Beaumont, M., Zhang, W., and Balding, D. (2002).
\newblock Approximate {B}ayesian computation in population genetics.
\newblock {\em Genetics}, 162(4):2025--2035.

\bibitem[Beaumont, 2019]{Beaumont2019}
Beaumont, M.~A. (2019).
\newblock Approximate {B}ayesian computation.
\newblock {\em Annual Review of Statistics and Its Application}, 6(1):379--403.

\bibitem[Berger, 1985]{berger:1985}
Berger, J. (1985).
\newblock {\em Statistical Decision Theory and {B}ayesian Analysis}.
\newblock Springer-Verlag, New York, second edition.

\bibitem[Berild \emph{et~al.}, 2021]{berild2021importance}
Berild, M.~O., Martino, S., Gómez-Rubio, V., and Rue, H. (2021).
\newblock Importance sampling with the integrated nested {L}aplace
  approximation.

\bibitem[Bernton \emph{et~al.}, 2019]{Bernton2019}
Bernton, E., Jacob, P.~E., Gerber, M., and Robert, C.~P. (2019).
\newblock Approximate {B}ayesian computation with the {W}asserstein distance.
\newblock {\em J. Royal Statist. Society Series B}, 81(2):235--269.

\bibitem[Besag and Green, 1993]{besag:green:1993}
Besag, J. and Green, P. (1993).
\newblock Spatial statistics and {B}ayesian computation.
\newblock {\em J. Royal Statist. Society Series B}, 55(1):25--37.
\newblock With discussion.

\bibitem[{Betancourt}, 2018]{betancourt:2018}
{Betancourt}, M. (2018).
\newblock A conceptual introduction to {H}amiltonian {M}onte {C}arlo.
\newblock {\em https://arxiv.org/abs/1701.02434v2}.

\bibitem[Bilodeau \emph{et~al.}, 2021]{bilodeau2021stochastic}
Bilodeau, B., Stringer, A., and Tang, Y. (2021).
\newblock Stochastic convergence rates and applications of adaptive quadrature
  in {B}ayesian inference.

\bibitem[Bishop, 2006]{bishop2006pattern}
Bishop, C.~M. (2006).
\newblock {\em Pattern Recognition and Machine Learning}.
\newblock Springer, New York.

\bibitem[Bissiri \emph{et~al.}, 2016]{bissiri:etal:2016}
Bissiri, P.~G., Holmes, C.~C., and Walker, S.~G. (2016).
\newblock A general framework for updating belief distributions.
\newblock {\em J. Royal Statist. Society Series B}, 78(5):1103--1130.

\bibitem[Blei \emph{et~al.}, 2017]{blei2017variational}
Blei, D.~M., Kucukelbir, A., and McAuliffe, J.~D. (2017).
\newblock Variational inference: A review for statisticians.
\newblock {\em J. American Statist. Assoc.}, 112(518):859--877.

\bibitem[Blum, 2010]{blum:2010}
Blum, M. (2010).
\newblock {A}pproximate {B}ayesian computation: a non-parametric perspective.
\newblock {\em J. American Statist. Assoc.}, 105(491):1178--1187.

\bibitem[Blum and Fran{\c c}ois, 2010]{blum:francois:2010}
Blum, M. and Fran{\c c}ois, O. (2010).
\newblock Non-linear regression models for approximate {B}ayesian computation.
\newblock {\em Statist. Comput.}, 20:63--73.

\bibitem[Blum, 2017]{blum2019regression}
Blum, M.~G. (2017).
\newblock Regression approaches for approximate {B}ayesian computation.
\newblock {\em arXiv preprint arXiv:1707.01254}.

\bibitem[Blum \emph{et~al.}, 2013]{blum:etal:2013}
Blum, M. G.~B., Nunes, M.~A., Prangle, D., and Sisson, S.~A. (2013).
\newblock A comparative review of dimension reduction methods in approximate
  {B}ayesian computation.
\newblock {\em Statist. Science}, 28(2):189--208.

\bibitem[Bornn \emph{et~al.}, 2017]{bornn2017use}
Bornn, L., Pillai, N.~S., Smith, A., and Woodard, D. (2017).
\newblock The use of a single pseudo-sample in approximate {B}ayesian
  computation.
\newblock {\em Statist. Comp.}, 27(3):583--590.

\bibitem[Bortot \emph{et~al.}, 2007]{bordot2007}
Bortot, P., Coles, S.~G., and Sisson, S.~A. (2007).
\newblock Inference for stereological extremes.
\newblock {\em Journal of the American Statistical Association},
  102(477):84--92.

\bibitem[Braun and McAuliffe, 2010]{braun2010variational}
Braun, M. and McAuliffe, J. (2010).
\newblock Variational inference for large-scale models of discrete choice.
\newblock {\em J. American Statist. Assoc.}, 105(489):324--335.

\bibitem[Brooks \emph{et~al.}, 2011]{brooks:etal:2011}
Brooks, S., Gelman, A., Jones, G., and Meng, X. (2011).
\newblock {\em Handbook of {M}arkov {C}hain {M}onte {C}arlo}.
\newblock Taylor \& Francis.

\bibitem[Calvet and Czellar, 2015]{calvet2015accurate}
Calvet, L.~E. and Czellar, V. (2015).
\newblock Accurate methods for approximate {B}ayesian computation filtering.
\newblock {\em J. Finan. Econometrics}, 13(4):798--838.

\bibitem[Canale and Ruggiero, 2016]{canale2016}
Canale, A. and Ruggiero, M. (2016).
\newblock {B}ayesian nonparametric forecasting of monotonic functional time
  series.
\newblock {\em Electronic Journal of Statistics}, 10(2):3265--3286.

\bibitem[Carpenter \emph{et~al.}, 2017]{carpenter2017stan}
Carpenter, B., Gelman, A., Hoffman, M.~D., Lee, D., Goodrich, B., Betancourt,
  M., Brubaker, M., Guo, J., Li, P., and Riddell, A. (2017).
\newblock Stan: A probabilistic programming language.
\newblock {\em Journal of statistical software}, 76(1):1--32.

\bibitem[Casella and George, 1992]{casella:george:1992}
Casella, G. and George, E. (1992).
\newblock An introduction to {G}ibbs sampling.
\newblock {\em American Statist.}, 46:167--174.

\bibitem[Ceruzzi, 2003]{ceruzzi:2003}
Ceruzzi, P. (2003).
\newblock {\em A History of Modern Computing}.
\newblock MIT Press, second edition.

\bibitem[Chan and Yu, 2020]{chan2020fast}
Chan, J.~C. and Yu, X. (2020).
\newblock Fast and accurate variational inference for large {B}ayesian {VAR}s
  with stochastic volatility.
\newblock {\em CAMA Working Paper}.

\bibitem[Chen \emph{et~al.}, 2011]{chen2011}
Chen, S., Dick, J., and Owen, A.~B. (2011).
\newblock Consistency of {M}arkov chain quasi-{M}onte {C}arlo on continuous
  state spaces.
\newblock {\em Ann. Statist.}, 39(2):673--701.

\bibitem[Chib, 2011]{chib2011introduction}
Chib, S. (2011).
\newblock Introduction to simulation and {MCMC} methods.
\newblock {\em The Oxford Handbook of {B}ayesian Econometrics}, pages 183--217.
\newblock OUP. Eds. Geweke, J., Koop, G. and van Dijk, H.

\bibitem[Chib and Greenberg, 1995]{chib:greenberg:1995}
Chib, S. and Greenberg, E. (1995).
\newblock Understanding the {M}etropolis--{H}astings algorithm.
\newblock {\em American Statist.}, 49:327--335.

\bibitem[Chib and Greenberg, 1996]{chib_greenberg_1996}
Chib, S. and Greenberg, E. (1996).
\newblock Markov chain {M}onte {C}arlo simulation methods in econometrics.
\newblock {\em Econometric Theory}, 12(3):409–431.

\bibitem[Chib \emph{et~al.}, 2018]{chib2018bayesian}
Chib, S., Shin, M., and Simoni, A. (2018).
\newblock {B}ayesian estimation and comparison of moment condition models.
\newblock {\em J. American Statist. Assoc.}, 113(524):1656--1668.

\bibitem[Clarté \emph{et~al.}, 2020]{clarte2019component}
Clarté, G., Robert, C.~P., Ryder, R.~J., and Stoehr, J. (2020).
\newblock {Componentwise approximate Bayesian computation via Gibbs-like
  steps}.
\newblock {\em Biometrika}, 108(3):591--607.

\bibitem[Davis and Rabinowitz, 1975]{davis1975numerical}
Davis, P. and Rabinowitz, P. (1975).
\newblock {\em Numerical Methods of Integration}.
\newblock Academic Press, New York.

\bibitem[Dean \emph{et~al.}, 2014]{dean2014parameter}
Dean, T.~A., Singh, S.~S., Jasra, A., and Peters, G.~W. (2014).
\newblock Parameter estimation for hidden {M}arkov models with intractable
  likelihoods.
\newblock {\em Scandinavian Journal of Statistics}, 41(4):970--987.

\bibitem[Deligiannidis \emph{et~al.}, 2018]{deligiannidis2018correlated}
Deligiannidis, G., Doucet, A., and Pitt, M.~K. (2018).
\newblock The correlated pseudomarginal method.
\newblock {\em J. Royal Statist. Society Series B}, 80(5):839--870.

\bibitem[Devroye, 1986]{devroye:1986}
Devroye, L. (1986).
\newblock {\em Non-Uniform Random Variate Generation}.
\newblock Springer-Verlag, New York.

\bibitem[Doucet \emph{et~al.}, 2015]{doucet2015efficient}
Doucet, A., Pitt, M.~K., Deligiannidis, G., and Kohn, R. (2015).
\newblock Efficient implementation of {M}arkov chain {M}onte {C}arlo when using
  an unbiased likelihood estimator.
\newblock {\em Biometrika}, 102(2):295--313.

\bibitem[Drovandi and Frazier, 2021]{drovandi2021comparison}
Drovandi, C. and Frazier, D.~T. (2021).
\newblock A comparison of likelihood-free methods with and without summary
  statistics.
\newblock {\em arXiv preprint arXiv:2103.02407}.

\bibitem[Drovandi \emph{et~al.}, 2011]{drovandi:pettitt:faddy:2011}
Drovandi, C., Pettitt, A., and Faddy, M. (2011).
\newblock Approximate {B}ayesian computation using indirect inference.
\newblock {\em J. Royal Statist. Society Series A}, 60(3):503--524.

\bibitem[Drovandi \emph{et~al.}, 2015]{drovandi2015bayesian}
Drovandi, C.~C., Pettitt, A.~N., and Lee, A. (2015).
\newblock {B}ayesian indirect inference using a parametric auxiliary model.
\newblock {\em Statist. Science}, 30(1):72--95.

\bibitem[Dunson and Johndrow, 2019]{dunson2019hastings}
Dunson, D. and Johndrow, J. (2019).
\newblock The {H}astings algorithm at fifty.
\newblock {\em Biometrika}, 107(1):1--23.

\bibitem[Elvira and Martino, 2021]{elvira2021advances}
Elvira, V. and Martino, L. (2021).
\newblock Advances in importance sampling.

\bibitem[Fearnhead, 2018]{fearnhead2018asymptotics}
Fearnhead, P. (2018).
\newblock Asymptotics of {ABC}.
\newblock {\em Handbook of Approximate {B}ayesian Computation}, pages 269--288.
\newblock Chapman \& Hall/CRC. Eds. Sisson, S., Fan, Y., Beaumont, M.

\bibitem[Fearnhead and Prangle, 2012]{fearnhead:prangle:2012}
Fearnhead, P. and Prangle, D. (2012).
\newblock Constructing summary statistics for approximate {B}ayesian
  computation: {S}emi-automatic approximate {B}ayesian computation.
\newblock {\em J. Royal Statist. Society Series B}, 74(3):419--474.
\newblock With discussion.

\bibitem[Fienberg, 2006]{fienberg:2006}
Fienberg, S. (2006).
\newblock When did {B}ayesian inference become ``{B}ayesian``?
\newblock {\em {B}ayesian Analysis}, 1(1):1--40.

\bibitem[Frazier, 2020]{frazier2020robust}
Frazier, D.~T. (2020).
\newblock Robust and efficient {A}pproximate {B}ayesian {C}omputation: A
  minimum distance approach.
\newblock {\em arXiv preprint arXiv:2006.14126}.

\bibitem[Frazier and Drovandi, 2019]{frazier2019robust}
Frazier, D.~T. and Drovandi, C. (2019).
\newblock Robust approximate {B}ayesian inference with synthetic likelihood.
\newblock {\em https://arXiv:1904.04551}.

\bibitem[Frazier \emph{et~al.}, 2021a]{frazier2021synthetic}
Frazier, D.~T., Drovandi, C., and Nott, D.~J. (2021a).
\newblock Synthetic likelihood in misspecified models: Consequences and
  corrections.
\newblock {\em arXiv preprint arXiv:2104.03436}.

\bibitem[Frazier \emph{et~al.}, 2021b]{frazier2021note}
Frazier, D.~T., Loaiza-Maya, R., and Martin, G.~M. (2021b).
\newblock A note on the accuracy of variational {B}ayes in state space models:
  Inference and prediction.

\bibitem[Frazier \emph{et~al.}, 2021c]{frazierloss}
Frazier, D.~T., Loaiza-Maya, R., Martin, G.~M., and Koo, B. (2021c).
\newblock Loss-based variational {B}ayes prediction.
\newblock {\em arXiv preprint arXiv:2104.14054}.

\bibitem[Frazier \emph{et~al.}, 2019a]{frazier2019approximate}
Frazier, D.~T., Maneesoonthorn, W., Martin, G.~M., and McCabe, B.~P. (2019a).
\newblock Approximate {B}ayesian forecasting.
\newblock {\em Intern. J. Forecasting}, 35(2):521--539.

\bibitem[Frazier \emph{et~al.}, 2018]{FMRR2016}
Frazier, D.~T., Martin, G.~M., Robert, C.~P., and Rousseau, J. (2018).
\newblock Asymptotic properties of approximate {B}ayesian computation.
\newblock {\em Biometrika}, 105(3):593--607.

\bibitem[Frazier \emph{et~al.}, 2019b]{frazier2019bayesian}
Frazier, D.~T., Nott, D.~J., Drovandi, C., and Kohn, R. (2019b).
\newblock {B}ayesian inference using synthetic likelihood: {A}symptotics and
  adjustments.
\newblock {\em https://arXiv:1902.04827}.

\bibitem[Frazier \emph{et~al.}, 2020]{frazier2020model}
Frazier, D.~T., Robert, C.~P., and Rousseau, J. (2020).
\newblock Model misspecification in approximate {B}ayesian computation:
  consequences and diagnostics.
\newblock {\em J. Royal Statist. Society Series B}.

\bibitem[Gallant and Tauchen, 1996]{gallant1996moments}
Gallant, A.~R. and Tauchen, G. (1996).
\newblock Which moments to match?
\newblock {\em Econometric theory}, 12(4):657--681.

\bibitem[Gelfand and Smith, 1990]{gelfand:smith90}
Gelfand, A. and Smith, A. (1990).
\newblock {S}ampling based approaches to calculating marginal densities.
\newblock {\em J. Amer. Statist. Assoc.}, 85(410):398--409.

\bibitem[Geman and Geman, 1984]{geman:1984}
Geman, S. and Geman, D. (1984).
\newblock Stochastic relaxation, {G}ibbs distributions and the {B}ayesian
  restoration of images.
\newblock {\em IEEE Trans. Pattern Anal. Mach. Intell.}, 6:721--741.

\bibitem[Gerber and Chopin, 2015]{gerber:chopin:2015}
Gerber, M. and Chopin, N. (2015).
\newblock Sequential quasi {M}onte {C}arlo.
\newblock {\em J. Royal Statist. Society Series B}, 77(3):509--579.

\bibitem[Geweke, 1989]{geweke:1989}
Geweke, J. (1989).
\newblock {B}ayesian inference in econometric models using {M}onte {C}arlo
  integration.
\newblock {\em Econometrica}, 57(6):1317--1340.

\bibitem[Geweke \emph{et~al.}, 2011]{geweke2011handbook}
Geweke, J., Koop, G., and van Dijk, H. (2011).
\newblock {\em The Oxford Handbook of {B}ayesian Econometrics}.
\newblock OUP.

\bibitem[Geyer, 2011]{geyer2011introduction}
Geyer, C.~J. (2011).
\newblock Introduction to {M}arkov chain {M}onte {C}arlo.
\newblock {\em Handbook of Markov chain Monte Carlo}, pages 3--48.
\newblock Chapman \& Hall/CRC. Eds. Brooks, S., Gelman, A., Jones, G., Meng,
  X-L.

\bibitem[Gomez-Rubio and Rue, 2018]{Gomez2018}
Gomez-Rubio, V. and Rue, H. (2018).
\newblock Markov chain {M}onte {C}arlo with the integrated nested {L}aplace
  approximation.
\newblock {\em Statistics and Computing}, 28(5).

\bibitem[Goodfellow \emph{et~al.}, 2014]{goodfellow2014generative}
Goodfellow, I., Pouget-Abadie, J., Mirza, M., Xu, B., Warde-Farley, D., Ozair,
  S., Courville, A., and Bengio, Y. (2014).
\newblock Generative adversarial nets.
\newblock In {\em Advances in Neural Information Processing Systems}, pages
  2672--2680.

\bibitem[Gordon \emph{et~al.}, 1993]{gordon:salmon:smith:1993}
Gordon, N., Salmond, J., and Smith, A. (1993).
\newblock A novel approach to non-linear/non-{G}aussian {B}ayesian state
  estimation.
\newblock {\em IEEE Proceedings on Radar and Signal Processing},
  140(2):107--113.

\bibitem[Gouri{\'e}roux \emph{et~al.}, 1993]{gourieroux:monfort:renault:1993}
Gouri{\'e}roux, C., Monfort, A., and Renault, E. (1993).
\newblock Indirect inference.
\newblock {\em J. Applied Econometrics}, 8:85--118.

\bibitem[Green \emph{et~al.}, 2015]{greenetal2015}
Green, P., Latuszynski, K., Pereyra, M., and Robert, C. (2015).
\newblock {B}ayesian computation: a summary of the current state, and samples
  backwards and forwards.
\newblock {\em Statist. Comp.}, 25:835--862.

\bibitem[Gutmann and Corander, 2016]{gutmann2016bayesian}
Gutmann, M.~U. and Corander, J. (2016).
\newblock {B}ayesian optimization for likelihood-free inference of
  simulator-based statistical models.
\newblock {\em The Journal of Machine Learning Research}, 17(1):4256--4302.

\bibitem[Hammersley and Handscomb, 1964]{hammersley:handscomb:1964}
Hammersley, J. and Handscomb, D. (1964).
\newblock {\em {M}onte {C}arlo Methods}.
\newblock John Wiley, New York.

\bibitem[Hastings, 1970]{hastings:1970}
Hastings, W. (1970).
\newblock {M}onte {C}arlo sampling methods using {M}arkov chains and their
  application.
\newblock {\em Biometrika}, 57(1):97--109.

\bibitem[Hoffman \emph{et~al.}, 2013]{hoffman13}
Hoffman, M.~D., Blei, D.~M., Wang, C., and Paisley, J. (2013).
\newblock Stochastic variational inference.
\newblock {\em Journal of Machine Learning Research}, 14(4):1303--1347.

\bibitem[Hoogerheide \emph{et~al.}, 2009]{hoogerheide2009simulation}
Hoogerheide, L.~F., van Dijk, H.~K., and van Oest, R.~D. (2009).
\newblock Simulation based {B}ayesian econometric inference: principles and
  some recent computational advances.
\newblock {\em Handbook of Computational Econometrics}, pages 215--280.
\newblock John Wiley \& Sons. Eds. van Dijk, H. and van Oest, R.

\bibitem[Huggins \emph{et~al.}, 2019]{huggins2019validated}
Huggins, J.~H., Kasprzak, M., Campbell, T., and Broderick, T. (2019).
\newblock Validated variational inference via practical posterior error bounds.
\newblock {\em https://arXiv:1910.04102}.

\bibitem[Jasra, 2015]{jasra2015approximate}
Jasra, A. (2015).
\newblock Approximate {B}ayesian computation for a class of time series models.
\newblock {\em International Statistical Review}, 83(3):405--435.

\bibitem[Jasra \emph{et~al.}, 2012]{jasra:etal:2012}
Jasra, A., Singh, S., Martin, J., and McCoy, E. (2012).
\newblock Filtering via approximate {B}ayesian computation.
\newblock {\em Statist. Comp.}, 22:1223--1237.

\bibitem[Jennings and Madigan, 2017]{jennings2017astroabc}
Jennings, E. and Madigan, M. (2017).
\newblock Astro{ABC}: an approximate {B}ayesian computation sequential {M}onte
  {C}arlo sampler for cosmological parameter estimation.
\newblock {\em Astronomy and Computing}, 19:16--22.

\bibitem[Jiang, 2018]{jiang2018approximate}
Jiang, B. (2018).
\newblock Approximate {B}ayesian computation with {K}ullback-{L}eibler
  divergence as data discrepancy.
\newblock In {\em International Conference on Artificial Intelligence and
  Statistics}, pages 1711--1721. PMLR.

\bibitem[Jiang and Tanner, 2008]{jiang2008}
Jiang, W. and Tanner, M.~A. (2008).
\newblock Gibbs posterior for variable selection in high-dimensional
  classification and data-mining.
\newblock {\em Annals of Statistics}, 36(5):2207--2231.

\bibitem[Johndrow \emph{et~al.}, 2019]{johndrow2019mcmc}
Johndrow, J.~E., Smith, A., Pillai, N., and Dunson, D.~B. (2019).
\newblock {MCMC} for imbalanced categorical data.
\newblock {\em J. American Statist. Assoc.}, 114(527):1394--1403.

\bibitem[Joyce and Marjoram, 2008]{joyce:marjoram:2008}
Joyce, P. and Marjoram, P. (2008).
\newblock Approximately sufficient statistics and {B}ayesian computation.
\newblock {\em Statistical Applications in Genetics and Molecular Biology},
  7(1):article 26.

\bibitem[Kabisa \emph{et~al.}, 2016]{kabisa2016online}
Kabisa, S., Dunson, D.~B., and Morris, J.~S. (2016).
\newblock Online variational {B}ayes inference for high-dimensional correlated
  data.
\newblock {\em J. Comput. Graph. Statist.}, 25(2):426--444.

\bibitem[Kloek and van Dijk, 1978]{kloek1978bayesian}
Kloek, T. and van Dijk, H.~K. (1978).
\newblock {B}ayesian estimates of equation system parameters: an application of
  integration by {M}onte {C}arlo.
\newblock {\em Econometrica}, 46(1):1--19.

\bibitem[Knoblauch \emph{et~al.}, 2019]{knoblauch2019generalized}
Knoblauch, J., Jewson, J., and Damoulas, T. (2019).
\newblock Generalized variational inference.
\newblock {\em https://arXiv:1904.02063}.

\bibitem[Kon Kam~King \emph{et~al.}, 2019]{konkamking2019}
Kon Kam~King, G., Canale, A., and Ruggiero, M. (2019).
\newblock {B}ayesian functional forecasting with locally-autoregressive
  dependent processes.
\newblock {\em {B}ayesian Anal.}, 14(4):1121--1141.

\bibitem[Koop and Korobilis, 2018]{koop2018variational}
Koop, G. and Korobilis, D. (2018).
\newblock Variational {B}ayes inference in high-dimensional time-varying
  parameter models.
\newblock {\em SSRN 3246472}.

\bibitem[Kousathanas \emph{et~al.}, 2019]{kousathanas2018guide}
Kousathanas, A., Duchen, P., and Wegmann, D. (2019).
\newblock A guide to general-purpose {ABC} software.
\newblock In {\em Handbook of approximate Bayesian computation}, pages
  369--413. Chapman and Hall/CRC.

\bibitem[Kousathanas \emph{et~al.}, 2016]{Kousathanas2016}
Kousathanas, A., Leuenberger, C., Helfer, J., Quinodoz, M., Foll, M., and
  Wegmann, D. (2016).
\newblock Likelihood-free inference in high-dimensional models.
\newblock {\em Genetics}, 203(2):893--904.

\bibitem[Kucukelbir \emph{et~al.}, 2017]{kucukelbir2017automatic}
Kucukelbir, A., Tran, D., Ranganath, R., Gelman, A., and Blei, D.~M. (2017).
\newblock Automatic differentiation variational inference.
\newblock {\em The Journal of Machine Learning Research}, 18(1):430--474.

\bibitem[Lemieux, 2009]{lemieux2009monte}
Lemieux, C. (2009).
\newblock {\em Monte {C}arlo and quasi-{M}onte {C}arlo sampling}.
\newblock Springer Science \& Business Media.

\bibitem[Li and Fearnhead, 2018a]{LF2016b}
Li, W. and Fearnhead, P. (2018a).
\newblock {C}onvergence of regression-adjusted approximate {B}ayesian
  computation.
\newblock {\em Biometrika}, 105(2):301--318.

\bibitem[Li and Fearnhead, 2018b]{LF2016a}
Li, W. and Fearnhead, P. (2018b).
\newblock {O}n the asymptotic efficiency of approximate {B}ayesian computation
  estimators.
\newblock {\em Biometrika}, 105(2):285--299.

\bibitem[Lintusaari \emph{et~al.}, 2017]{lintusaari2017fundamentals}
Lintusaari, J., Gutmann, M.~U., Dutta, R., Kaski, S., and Corander, J. (2017).
\newblock Fundamentals and recent developments in approximate {B}ayesian
  computation.
\newblock {\em Systematic biology}, 66(1):e66--e82.

\bibitem[Llorente \emph{et~al.}, 2021]{llorente2021marginal}
Llorente, F., Martino, L., Delgado, D., and Lopez-Santiago, J. (2021).
\newblock Marginal likelihood computation for model selection and hypothesis
  testing: an extensive review.
\newblock {\em https://arXiv:2005.08334}.

\bibitem[Loaiza-Maya \emph{et~al.}, 2021a]{loaiza2019focused}
Loaiza-Maya, R., Martin, G.~M., and Frazier, D.~T. (2021a).
\newblock Focused {B}ayesian prediction.
\newblock {\em Journal of Applied Econometrics}, 36(5):517--543.

\bibitem[Loaiza-Maya \emph{et~al.}, 2021b]{loaiza2020fast}
Loaiza-Maya, R., Smith, M.~S., Nott, D.~J., and Danaher, P.~J. (2021b).
\newblock Fast and accurate variational inference for models with many latent
  variables.
\newblock {\em Journal of Econometrics}.

\bibitem[Margossian \emph{et~al.}, 2020]{margossian2020}
Margossian, C.~C., Vehtari, A., Simpson, D., and Agrawal, R. (2020).
\newblock Hamiltonian {M}onte {C}arlo using an adjoint-differentiated {L}aplace
  approximation: Bayesian inference for latent {G}aussian models and beyond.
\newblock {\em https://arXiv:2004.12550}.

\bibitem[Marin \emph{et~al.}, 2011]{marin:pudlo:robert:ryder:2011}
Marin, J., Pudlo, P., Robert, C., and Ryder, R. (2011).
\newblock Approximate {B}ayesian computational methods.
\newblock {\em Statist. Comp.}, 21(2):279--291.

\bibitem[Marjoram \emph{et~al.}, 2003]{marjoram:etal:2003}
Marjoram, P., Molitor, J., Plagnol, V., and Tavar{\'e}, S. (2003).
\newblock Markov chain {M}onte {C}arlo without likelihoods.
\newblock {\em Proc.~Natl.~Acad.~Sci.~USA}, 100(26):15324--15328.

\bibitem[Martin \emph{et~al.}, 2020]{martin2020computing}
Martin, G.~M., Frazier, D.~T., and Robert, C.~P. (2020).
\newblock Computing {B}ayes: Bayesian computation from 1763 to the 21st
  century.
\newblock {\em https://arXiv:2004.06425}.

\bibitem[Martin \emph{et~al.}, 2019]{martin2019auxiliary}
Martin, G.~M., McCabe, B.~P., Frazier, D.~T., Maneesoonthorn, W., and Robert,
  C.~P. (2019).
\newblock Auxiliary likelihood-based approximate {B}ayesian computation in
  state space models.
\newblock {\em J. Comput. Graph. Statist.}, 28(3):508--522.

\bibitem[Martino and Riebler, 2019]{martino2019integrated}
Martino, S. and Riebler, A. (2019).
\newblock Integrated nested {L}aplace approximations ({INLA}).
\newblock {\em https://arXiv:1907.01248}.

\bibitem[Metropolis \emph{et~al.}, 1953]{metropolis:1953}
Metropolis, N., Rosenbluth, A.~W., Rosenbluth, M.~N., Teller, A.~H., and
  Teller, E. (1953).
\newblock Equations of state calculations by fast computing machines.
\newblock {\em J. Chem. Phys.}, 21:1087--1092.

\bibitem[Metropolis and Ulam, 1949]{metropolis:ulam:1949}
Metropolis, N. and Ulam, S. (1949).
\newblock The {M}onte {C}arlo method.
\newblock {\em J. American Statist. Assoc.}, 44:335--341.

\bibitem[Miller and Dunson, 2019]{miller2019robust}
Miller, J.~W. and Dunson, D.~B. (2019).
\newblock Robust {B}ayesian inference via coarsening.
\newblock {\em J. American Statist. Assoc.}, 114(527):1113--1125.

\bibitem[Moores \emph{et~al.}, 2020]{moores2020bayesian}
Moores, M.~T., Pettitt, A.~N., and Mengersen, K. (2020).
\newblock Bayesian computation with intractable likelihoods.

\bibitem[Naesseth \emph{et~al.}, 2019]{naesseth2019elements}
Naesseth, C.~A., Lindsten, F., Sch{\"o}n, T.~B.,\emph{et~al.} (2019).
\newblock Elements of sequential {M}onte {C}arlo.
\newblock {\em Foundations and Trends in Machine Learning}, 12(3):307--392.

\bibitem[Naylor and Smith, 1982]{naylor:smith:1982}
Naylor, J. and Smith, A. (1982).
\newblock Application of a method for the efficient computation of posterior
  distributions.
\newblock {\em Applied {S}tatistics}, 31(3):214--225.

\bibitem[Nguyen \emph{et~al.}, 2020]{nguyen2020approximate}
Nguyen, H.~D., Arbel, J., L{\"u}, H., and Forbes, F. (2020).
\newblock Approximate {B}ayesian computation via the energy statistic.
\newblock {\em IEEE Access}, 8:131683--131698.

\bibitem[Nott \emph{et~al.}, 2018]{nott2018high}
Nott, D., Ong, V. M.-H., Fan, Y., and Sisson, S. (2018).
\newblock High-dimensional {ABC}.
\newblock {\em Handbook of Approximate {B}ayesian Computation}, pages 211--242.
\newblock Chapman \& Hall/CRC. Eds. Sisson, S., Fan, Y., Beaumont, M.

\bibitem[Nott \emph{et~al.}, 2014]{Nott2014}
Nott, D.~J., Fan, Y., Marshall, L., and Sisson, S.~A. (2014).
\newblock Approximate {B}ayesian computation and {B}ayes’ linear analysis:
  Toward high-dimensional {ABC}.
\newblock {\em Journal of Computational and Graphical Statistics},
  23(1):65--86.

\bibitem[O'Hagan and West, 2010]{ohagan2010handbook}
O'Hagan, A. and West, M. (2010).
\newblock {\em The Oxford Handbook of Applied {B}ayesian Analysis}.
\newblock OUP.

\bibitem[Ong \emph{et~al.}, 2018]{ong2018variational}
Ong, V.~M., Nott, D.~J., Tran, M.-N., Sisson, S.~A., and Drovandi, C.~C.
  (2018).
\newblock Variational {B}ayes with synthetic likelihood.
\newblock {\em Statist. Comp.}, 28(4):971--988.

\bibitem[Ormerod and Wand, 2010]{ormerod2010explaining}
Ormerod, J.~T. and Wand, M.~P. (2010).
\newblock Explaining variational approximations.
\newblock {\em American Statist.}, 64(2):140--153.

\bibitem[Pacchiardi and Dutta, 2021]{pacchiardi2021generalized}
Pacchiardi, L. and Dutta, R. (2021).
\newblock Generalized bayesian likelihood-free inference using scoring rules
  estimators.

\bibitem[Park \emph{et~al.}, 2016]{park2016k2}
Park, M., Jitkrittum, W., and Sejdinovic, D. (2016).
\newblock K2-{ABC}: Approximate {B}ayesian computation with kernel embeddings.
\newblock In {\em Artificial Intelligence and Statistics}, pages 398--407.
  PMLR.

\bibitem[Peters \emph{et~al.}, 2012]{peters2012likelihood}
Peters, G.~W., Sisson, S.~A., and Fan, Y. (2012).
\newblock Likelihood-free {B}ayesian inference for $\alpha$-stable models.
\newblock {\em Comput. Statist. Data Anal.}, 56(11):3743--3756.

\bibitem[Price \emph{et~al.}, 2018]{price2018bayesian}
Price, L.~F., Drovandi, C.~C., Lee, A., and Nott, D.~J. (2018).
\newblock {B}ayesian synthetic likelihood.
\newblock {\em J. Comput. Graph. Statist.}, 27(1):1--11.

\bibitem[Pritchard \emph{et~al.}, 1999]{pritchard:seielstad:perez:feldman:1999}
Pritchard, J., Seielstad, M., Perez-Lezaun, A., and Feldman, M. (1999).
\newblock Population growth of human {Y} chromosomes: a study of {Y} chromosome
  microsatellites.
\newblock {\em Mol. Biol. Evol.}, 16:1791--1798.

\bibitem[Quiroz \emph{et~al.}, 2019]{quiroz2019speeding}
Quiroz, M., Kohn, R., Villani, M., and Tran, M.-N. (2019).
\newblock Speeding up {MCMC} by efficient data subsampling.
\newblock {\em J. American Statist. Assoc.}, 114(526):831--843.

\bibitem[Quiroz \emph{et~al.}, 2018a]{quiroz2018gaussian}
Quiroz, M., Nott, D.~J., and Kohn, R. (2018a).
\newblock Gaussian variational approximation for high-dimensional state space
  models.
\newblock {\em https://arXiv:1801.07873}.

\bibitem[Quiroz \emph{et~al.}, 2018b]{quiroz2018speeding}
Quiroz, M., Tran, M.-N., Villani, M., and Kohn, R. (2018b).
\newblock Speeding up {MCMC} by delayed acceptance and data subsampling.
\newblock {\em J. Comput. Graph. Statist.}, 27(1):12--22.

\bibitem[{R Core Team}, 2020]{Rlang}
{R Core Team} (2020).
\newblock {\em R: A Language and Environment for Statistical Computing}.
\newblock R Foundation for Statistical Computing, Vienna, Austria.

\bibitem[Robert, 2001]{robert:2001}
Robert, C. (2001).
\newblock {\em The {B}ayesian Choice}.
\newblock Springer-Verlag, New York, second edition.

\bibitem[Robert and Casella, 2011]{robert:casella:2011}
Robert, C. and Casella, G. (2011).
\newblock A history of {M}arkov chain {M}onte {C}arlo---subjective
  recollections from incomplete data.
\newblock {\em Statist. Science}, 26(1):102--115.

\bibitem[Robert \emph{et~al.}, 2018]{robert2018accelerating}
Robert, C.~P., Elvira, V., Tawn, N., and Wu, C. (2018).
\newblock Accelerating {MCMC} algorithms.
\newblock {\em Wiley Interdisciplinary Reviews: Computational Statistics},
  10(5):e1435.

\bibitem[Rodrigues \emph{et~al.}, 2019]{rodrigues2019likelihood}
Rodrigues, G., Nott, D.~J., and Sisson, S. (2019).
\newblock Likelihood-free approximate {G}ibbs sampling.
\newblock {\em https://arXiv:1906.04347}.

\bibitem[Rossi and Allenby, 2003]{rossi2003}
Rossi, P.~E. and Allenby, G.~M. (2003).
\newblock Bayesian statistics and marketing.
\newblock {\em Marketing Science}, 22(3):304--328.

\bibitem[Rue and Held, 2005]{rue:held:2005}
Rue, H. and Held, L. (2005).
\newblock {\em Gaussian {M}arkov Random Fields: {T}heory and Applications},
  volume 104 of {\em Monographs on Statistics and Applied Probability}.
\newblock Chapman \& Hall, London.

\bibitem[Rue \emph{et~al.}, 2009]{rue:martino:chopin:2009}
Rue, H., Martino, S., and Chopin, N. (2009).
\newblock Approximate {B}ayesian inference for latent {G}aussian models using
  integrated nested {L}aplace approximations.
\newblock {\em J. Royal Statist. Society Series B}, 71(2):319--392.

\bibitem[Rue \emph{et~al.}, 2017]{Rue2017}
Rue, H., Riebler, A., Sørbye, S.~H., Illian, J.~B., Simpson, D.~P., and
  Lindgren, F.~K. (2017).
\newblock Bayesian computing with inla: A review.
\newblock {\em Annual Review of Statistics and Its Application}, 4(1):395--421.

\bibitem[Schmon \emph{et~al.}, 2021]{schmon2021generalized}
Schmon, S.~M., Cannon, P.~W., and Knoblauch, J. (2021).
\newblock Generalized posteriors in approximate {B}ayesian computation.

\bibitem[Sisson and Fan, 2011]{sisson2011likelihood}
Sisson, S. and Fan, Y. (2011).
\newblock Likelihood-free {M}arkov chain {M}onte {C}arlo.
\newblock {\em Handbook of Markov Chain Monte Carlo}, pages 313--333.
\newblock Chapman \& Hall/CRC. Eds. Brooks, S., Gelman, A., Jones, G., Meng,
  X-L.

\bibitem[Sisson and Fan, 2019]{sissonfan2019}
Sisson, S. and Fan, Y. (2019).
\newblock {ABC} samplers.
\newblock {\em Handbook of Approximate {B}ayesian Computation}, pages 88--123.
\newblock Chapman \& Hall/CRC. Eds. Sisson, S., Fan, Y., Beaumont, M.

\bibitem[Sisson \emph{et~al.}, 2019]{sisson2018handbook}
Sisson, S.~A., Fan, Y., and Beaumont, M. (2019).
\newblock {\em Handbook of Approximate {B}ayesian Computation}.
\newblock Chapman \& Hall/CRC.

\bibitem[Sisson \emph{et~al.}, 2007]{sisson:fan:tanaka:2007}
Sisson, S.~A., Fan, Y., and Tanaka, M. (2007).
\newblock Sequential {M}onte {C}arlo without likelihoods.
\newblock {\em Proc. Natl. Acad. Sci. USA}, 104(6):1760--1765.

\bibitem[Smith and Roberts, 1993]{smith:roberts:1993}
Smith, A. and Roberts, G. (1993).
\newblock {B}ayesian computation via the {G}ibbs sampler and related {M}arkov
  chain {M}onte {C}arlo methods.
\newblock {\em J. Royal Statist. Society Series B}, 55(1):3--24.
\newblock With discussion.

\bibitem[Stigler, 1986a]{stigler:1986}
Stigler, S. (1986a).
\newblock {\em The History of {S}tatistics}.
\newblock Belknap, Cambridge.

\bibitem[Stigler, 1986b]{stigler:Laplace1774}
Stigler, S. (1986b).
\newblock Memoir on inverse probability.
\newblock {\em Statistical Science}, 1(3):359--363.

\bibitem[Stigler, 1975]{stigler:1975}
Stigler, S.~M. (1975).
\newblock Studies in the history of probability and statistics. {XXXIV}
  {N}apoleonic statistics: The work of {L}aplace.
\newblock {\em Biometrika}, 62(2):503--517.

\bibitem[Stoehr, 2017]{stoehr2017review}
Stoehr, J. (2017).
\newblock A review on statistical inference methods for discrete {M}arkov
  random fields.
\newblock {\em https://arXiv:1704.03331}.

\bibitem[Stringer \emph{et~al.}, 2021]{stringer2021}
Stringer, A., Brown, P., and Stafford, J. (2021).
\newblock Fast, scalable approximations to posterior distributions in extended
  latent {G}aussian models.

\bibitem[Tang and Reid, 2021]{tang2021}
Tang, Y. and Reid, N. (2021).
\newblock Laplace and saddlepoint approximations in high dimensions.

\bibitem[Tanner and Wong, 1987]{tanner87}
Tanner, M.~A. and Wong, W. (1987).
\newblock The calculation of posterior distributions by data augmentation.
\newblock {\em J. American Statist. Assoc.}, 82(398):528--550.
\newblock With discussion.

\bibitem[Tavar{\'e} \emph{et~al.}, 1997]{tavare:balding:griffith:donnelly:1997}
Tavar{\'e}, S., Balding, D., Griffith, R., and Donnelly, P. (1997).
\newblock Inferring coalescence times from {DNA} sequence data.
\newblock {\em Genetics}, 145:505--518.

\bibitem[Tierney and Kadane, 1986]{tierney:kadane:1986}
Tierney, L. and Kadane, J. (1986).
\newblock Accurate approximations for posterior moments and marginal densities.
\newblock {\em J. American Statist. Assoc.}, 81(393):82--86.

\bibitem[Tierney \emph{et~al.}, 1989]{tierney:kass:kadane:1989}
Tierney, L., Kass, R., and Kadane, J. (1989).
\newblock Fully exponential {L}aplace approximations to expectations and
  variances of non-positive functions.
\newblock {\em J. American Statist. Assoc.}, 84(407):710--716.

\bibitem[Tokdar and Kass, 2010]{Tokdar2010}
Tokdar, S. and Kass, R. (2010).
\newblock Importance sampling: A review.
\newblock {\em Wiley Interdisciplinary Reviews: Computational Statistics}, 2:54
  -- 60.

\bibitem[Tran \emph{et~al.}, 2019]{tran2019variational}
Tran, M.-N., Nguyen, D.~H., and Nguyen, D. (2019).
\newblock Variational {B}ayes on manifolds.
\newblock {\em https://arXiv:1908.03097}.

\bibitem[Tran \emph{et~al.}, 2017]{tran2017variational}
Tran, M.-N., Nott, D.~J., and Kohn, R. (2017).
\newblock Variational {B}ayes with intractable likelihood.
\newblock {\em J. Comput. Graph. Statist.}, 26(4):873--882.

\bibitem[Turner and Sederberg, 2014]{turner2014generalized}
Turner, B.~M. and Sederberg, P.~B. (2014).
\newblock A generalized, likelihood-free method for posterior estimation.
\newblock {\em Psychonomic bulletin \& review}, 21(2):227--250.

\bibitem[van Niekerk \emph{et~al.}, 2019]{vanniekerk2019new}
van Niekerk, J., Bakka, H., Rue, H., and Schenk, O. (2019).
\newblock New frontiers in {B}ayesian modeling using the {INLA} package in {R}.

\bibitem[Vanhatalo \emph{et~al.}, 2010]{Vanhatalo2010}
Vanhatalo, J., Pietiläinen, V., and Vehtari, A. (2010).
\newblock Approximate inference for disease mapping with sparse {G}aussian
  processes.
\newblock {\em Statistics in Medicine}, 29(15):1580--1607.

\bibitem[Vanslette \emph{et~al.}, 2019]{vanslette2019simple}
Vanslette, K., Alsheikh, A.~A., and Youcef-Toumi, K. (2019).
\newblock Why simple quadrature is just as good as {M}onte {C}arlo.
\newblock {\em https://arXiv:1908.00947}.

\bibitem[Wand, 2017]{wand2017fast}
Wand, M.~P. (2017).
\newblock Fast approximate inference for arbitrarily large semiparametric
  regression models via message passing.
\newblock {\em J. American Statist. Assoc.}, 112(517):137--168.

\bibitem[Wang and Blei, 2019a]{wangblei2019b}
Wang, Y. and Blei, D. (2019a).
\newblock Variational {B}ayes under model misspecification.
\newblock In {\em Advances in Neural Information Processing Systems}, pages
  13357--13367.

\bibitem[Wang and Blei, 2019b]{wangblei2019a}
Wang, Y. and Blei, D.~M. (2019b).
\newblock Frequentist consistency of variational {B}ayes.
\newblock {\em J. American Statist. Assoc.}, 114(527):1147--1161.

\bibitem[Wilkinson, 2013]{wilkinson:2013}
Wilkinson, R. (2013).
\newblock Approximate {B}ayesian computation {(ABC)} gives exact results under
  the assumption of model error.
\newblock {\em Statistical Applications in Genetics and Molecular Biology},
  12(2):129--141.

\bibitem[Wood, 2010]{wood:2010}
Wood, S. (2010).
\newblock Statistical inference for noisy nonlinear ecological dynamic systems.
\newblock {\em Nature}, 466(7310):1102--–1104.

\bibitem[Wood, 2019]{wood2019simplified}
Wood, S. (2019).
\newblock Simplified integrated nested {L}aplace approximation.
\newblock {\em Biometrika}, 107(1):223--230.

\bibitem[Yao \emph{et~al.}, 2018]{yao18}
Yao, Y., Vehtari, A., Simpson, D., and Gelman, A. (2018).
\newblock Yes, but did it work?: Evaluating variational inference.
\newblock {\em Proceedings of the 35th International Conference on Machine
  Learning}, 80:5581--5590.

\bibitem[Yu \emph{et~al.}, 2019]{yu2019assessment}
Yu, X., Nott, D.~J., Tran, M.-N., and Klein, N. (2019).
\newblock Assessment and adjustment of approximate inference algorithms using
  the law of total variance.
\newblock {\em https://arXiv:1911.08725}.

\bibitem[Zhang \emph{et~al.}, 2018]{zhang2018advances}
Zhang, C., B{\"u}tepage, J., Kjellstr{\"o}m, H., and Mandt, S. (2018).
\newblock Advances in variational inference.
\newblock {\em IEEE transactions on pattern analysis and machine intelligence},
  41(8):2008--2026.

\bibitem[Zhang and Gao, 2020]{zhang2017convergence}
Zhang, F. and Gao, C. (2020).
\newblock Convergence rates of variational posterior distributions.
\newblock {\em The Annals of Statistics}, 48(4):2180--2207.

\bibitem[Zhang, 2006a]{Zhang2006a}
Zhang, T. (2006a).
\newblock From eps-entropy to {KL} entropy: analysis of minimum information
  complexity density estimation.
\newblock {\em Annals of Statistics}, 34:2180–2210.

\bibitem[Zhang, 2006b]{Zhang2006b}
Zhang, T. (2006b).
\newblock Information-theoretic upper and lower bounds for statistical
  estimation.
\newblock {\em IEEE Trans. Info. Theory}, 52(4):1307–1321.

\end{thebibliography}
}

\end{document}